\documentclass[a4paper,11pt]{article}

\usepackage[margin=1in]{geometry}

\usepackage{amsthm}
\usepackage{amssymb}
\usepackage{amsmath}
\usepackage{tikz-cd}
\usepackage{array}
\usepackage{makecell}

\usepackage{hyperref}
\hypersetup{
    colorlinks=true,
    linkcolor=blue,
    filecolor=blue,      
    urlcolor=blue,
    citecolor = blue,
}
 
 \numberwithin{equation}{section}

\begin{document} 
{\center{\Huge Quantum mechanics in magnetic backgrounds with manifest symmetry and locality}\\~\\
{\Large Joe Davighi,$^1$ Ben Gripaios,$^2$ and Joseph Tooby-Smith$^2$}\\~\\
$^1$Department of Applied Mathematics and Theoretical Physics, University of Cambridge, Wilberforce Road, Cambridge, UK\\
$^2$Cavendish Laboratory, University of Cambridge, J.~J.~Thomson Ave, Cambridge, UK\\ ~\\
Emails: jed60@cam.ac.uk, gripaios@hep.phy.cam.ac.uk and jss85@cam.ac.uk}
~\\
~\\
~\\

\textbf{Abstract:} The usual methods for formulating and solving the quantum mechanics of a particle moving in a magnetic field respect neither locality nor any global symmetries which happen to be present. For example, Landau's solution for a particle moving in a uniform magnetic field in the plane involves choosing a gauge in which neither translation nor rotation invariance are manifest. We show that locality can be made manifest by passing to a redundant description in which the particle moves on a $U(1)$-principal bundle over the original configuration space and that symmetry can be made manifest by passing to a corresponding central extension of the original symmetry group by $U(1)$. With the symmetry manifest, one can attempt to solve the problem by using harmonic analysis and we provide a number of examples where this succeeds. One is a solution of the Landau problem in an arbitrary gauge (with either translation invariance or the full Euclidean group manifest). Another example is the motion of a fermionic rigid body, which can be formulated and solved in a manifestly local and symmetric way via a flat connection on the non-trivial $U(1)$-central extension of the configuration space $SO(3)$ given by $U(2)$.
\newpage
\tableofcontents

\section{Introduction}
Consider a particle moving on a smooth, connected, manifold $M$ in the presence of some background magnetic field. Suppose furthermore that the dynamics is invariant under some, connected, Lie group $G$ of global symmetries acting smoothly on $M$.

The study of the quantum mechanics of such a system is complicated by two well-known facts. The first complication is that it is, in general, not possible to write down a term in the lagrangian representing the magnetic field that is valid globally on $M$. Instead, the best that one can do is to cover $M$ by overlapping patches and to use 
multiple lagrangians, each of which is valid only locally on some patch. The most famous example, due to Dirac~\cite{dirac1931quantised} and solved by Tamm~\cite{tamm1931verallgemeinerten} (see also~\cite{wu1976dirac,WU1976365}), is given by the motion of an electrically-charged particle in the presence of a magnetic monopole, but we will see that there exists an example that is arguably even simpler (and certainly more prevalent in everyday life!), given by the motion of a rigid body which happens to be a fermion.

This latter example is interesting for another reason, which is that it shows that our set-up includes systems in which there is no apparent magnetic field, but rather a vector potential is being used to encode a global topological effect -- spin, in the case at hand -- in a manifestly local way. Thus, we will be able to write a local term in the lagrangian that accounts for the extra factor of $-1$ that the state of the fermion acquires when it undergoes a complete rotation, rather than arbitrarily assigning it by hand, as is usually done. This is desirable, given our prejudice that physics should be local.

The second complication is that the corresponding lagrangian (or lagrangians) will not be invariant under the action of $G$, but rather will shift by a total derivative. Perhaps the simplest example, made famous by Landau~\cite{Landau1930}, is given by the motion of a particle in a plane in the presence of a uniform magnetic field, where there is no choice of gauge such that the lagrangian is invariant under translations in more than one direction.

At the classical level, neither of these complications causes any problems, since they disappear once we pass from the lagrangian to the classical equations of motion. Indeed, the equations of motion are both globally valid and invariant (or rather covariant) under $G$. Thus, we can attempt to solve for the classical dynamics using our usual arsenal of techniques. But this is not the case at the quantum level. There, our usual technique is to convert the hamiltonian into an operator on $L^2(M)$ and to exploit the conserved charges corresponding to $G$ to solve, at least partially, the resulting Schr\"{o}dinger equation. Here though, we do not have a unique hamiltonian, but rather several; even if we did have a unique hamiltonian, we would, in general, find that the na\"{\i}ve operators corresponding to the conserved charges of $G$ do not commute with it. The last problem is often remedied by redefining the conserved charges, but then one finds that the new charges do not form a Lie algebra, unless we add further charges.

These two complications are apparently unrelated, at least as we have presented them. But they are related in the sense that neither could occur in the first place, were it not for a basic tenet of quantum mechanics, namely that physical states are represented by rays in a Hilbert space. Thus, the overall phase of a vector in a Hilbert space is not physical. 
This is what makes it possible, ultimately, to resolve the apparent paradox that, at a point in $M$ where two patches overlap, we have multiple, distinct lagrangians, but each of them gives rise to the same physics. Similarly, it allows us to absorb extra phases that arise from boundary contributions in the path integral under a $G$ transformation, when the lagrangian is not strictly invariant.

In this work we show that, by exploiting this basic property, one can formulate and solve (or at least, attempt to solve) such quantum systems in a unified way, using methods from harmonic analysis. In a nutshell, the
idea is as follows. A magnetic field defines a connection on a $U(1)$-principal bundle $P$ over $M$. From $G$ (which acts on $M$), we can construct a central extension $\tilde{G}$ of $G$ by $U(1)$ (which depends on the connection and on $P$, and which acts on $P$). We reformulate the original dynamical system on $M$ in terms of an equivalent system (with a redundant degree of freedom) of a particle moving on $P$. This reformulation allows us to circumvent both of the complications discussed above: not only do we have a unique, globally-valid, local lagrangian on $P$, but also the Hilbert space carries a {\em bona fide} representation of $\tilde{G}$ (in contrast to the original theory, in which the Hilbert space carries a projective representation of $G$, corresponding to the fact that a quantum state is represented by a ray in a Hilbert space). As a result, we can attempt a solution using harmonic analysis, with respect to the group $\tilde{G}$.

It should be remarked that neither the formulation nor the method of solution that we describe here can really be considered new. The formulation via central extensions has appeared in a number of places in the literature, mainly with applications to symplectic geometry and geometric quantization (see {\em e.g.}, \cite{marmo1988quasi, TUYNMAN1987207}) and the use of harmonic analysis to solve quantum systems in the absence of magnetic fields (and hence without the complications described above) was described in \cite{Gripaios2016}. What is new, we hope, is the synthesis of these ideas, which leads to a uniform approach to solving quantum-mechanical systems, including cases with magnetic fields (a type of topological interaction due to its independence from the worldvolume metric) or other non-trivial topological terms.

We remark in passing that our general formalism differs from that used in the study of integrable systems. In an integrable system one requires there exist a set of mutually commuting charges, while for us the charges are allowed to form any Lie algebra. Moreover, in our systems, the charges must correspond to the group action on the position space manifold. 
That said, it is worth noting that a number of the quantum mechanics models we consider turn out to be superintegrable, offering a complementary way of understanding their exact solvability. For instance, the Landau system is rendered maximally superintegrable by the fact that it is symmetric under the full Euclidean group in two dimensions, providing a set of three independent conserved charges (which we may take to be the Hamiltonian and the two Johnson-Lippmann charges~\cite{Johnson_Lippmann_1949}), two of which are in involution~\cite{McSween_Winternitz_2000,Berube_Winternitz_2004,Rodriguez}. We exploit this same basic fact in \S \ref{Sec_LandauISO} to solve the Landau system, but rather using a central extension of the 2d Euclidean group.

The methods we present are most powerful in cases where $G$ acts transitively on $M$ (meaning that any point in $M$ can be reached from any other via the action of $G$) corresponding to a special case ($0+1$ spacetime dimensions) of the usual non-linear sigma model of quantum field theory on a homogeneous space $G/H$. The constraint that $G$ acts transitively is a strong one; it implies, in particular, that any potential term in the lagrangian must be a constant. 
We thus have a `free' particle, in the sense that, in the absence of the magnetic field (and ignoring possible higher-derivative terms), the classical trajectories are given by the geodesics of some $G$-invariant metric. Despite the strong restrictions, one finds that a large class of interesting quantum mechanical models fall into this class and can be solved in this way. Examples discussed in the sequel include the systems considered by Landau (which, in contrast with Landau, we solve by keeping a transitive group of symmetries - either translations or the full Euclidean group - manifest) and Dirac (where we constrain the particle to move on the surface of a sphere, so that the rotation group acts transitively).

In cases where $G$ does not act transitively, the methods typically provide only a partial solution, in that they allow us to reduce the Schr\"{o}dinger equation to one on the space of orbits of $G$. But even here we find interesting examples where a complete solution is possible.

Since the existing literature underlying this work is somewhat arcane,
and since we hope that our results may be of interest to physicists and chemists who are not so mathematically inclined, we aim for a discussion that is both pedagogical and reasonably self-contained (in particular, pertinent mathematical definitions are supplied in Appendix \ref{sec_useful_mathematics}). Thus, we start by illustrating the ideas with elementary (but incomplete) discussions of the examples of planar motion in a uniform magnetic field (\S \ref{Sec_LandauLevels}) and of rigid body rotation (\S \ref{Sec_RigidBody}). These examples are particularly transparent because, for the former, the bundle is (topologically) trivial, so all the effects come from the magnetic field, while for the latter, the magnetic field vanishes (though the vector potential does not) so all effects arise from the topology of the bundle. 

After this, in \S\ref{Formalism}, we give full mathematical details of the method. We then complete the discussion of rigid body rotation (\S \ref{Rigid_Rotation}) and give a series of other examples which illustrate the method: the Dirac monopole (\S \ref{Sec_DiracMonopole}), a charged particle in the electromagnetic field of a dyon (\S \ref{Sec_dyon}), a repeat of Landau levels on a plane, but using the full Euclidean group (\S\ref{Sec_LandauISO}), motion on the Heisenberg group manifold (\S \ref{Sec_Heisenberg}), and motion in a uniform magnetic field with a mass that varies with position (\S \ref{Sec_varymass}), the last of which gives a completely solvable example in the case where the action of $G$ on $M$ is not transitive. All the examples considered in this paper are summarised in Table \ref{Example_table}.

In \S \ref{sec_nonTransv}, we discuss one further subtlety: it has long been known \cite{Manton1983,Manton1985} that only a subgroup of the symmetry of the classical equations of motion will be well-defined at the quantum level, so we discuss what happens in such cases. Such anomalies can occur in the presence of a magnetic background, dispite the absence of chiral fermions. Our conclusions are presented in \S\ref{Sec_Discussion}.

\section{Prototypes} \label{Sec_Prototyes}
\subsection{Planar motion in a uniform magnetic field} \label{Sec_LandauLevels}
Our first example is one made famous by Landau, in which a particle moves in the $xy$-plane with a uniform magnetic field $B\in\mathbb{R}$ in the $z$-direction. In this example, the subtleties are entirely due to the presence of the magnetic field. In particular, no matter what gauge is chosen, the usual lagrangian shifts by a non-vanishing total derivative under the action of the symmetry group, which for the purposes of the present discussion we take to be translations in $\mathbb{R}^2$. As a result, the usual quantum hamiltonian does not commute with the momenta and one cannot solve via a Fourier transform (which corresponds to harmonic analysis with respect to the group $\mathbb{R}^2$).

To circumvent this we write the action, contributing to the action phase $e^{iS}$, as
\begin{equation} \label{Action_LaH}
S=\int dt\left(\frac{1}{2} \dot x^2+\frac{1}{2} \dot y^2- \dot s- B y \dot x\right),
\end{equation}
with an additional degree of freedom $s\in \mathbb{R}$, with $s \sim s+2\pi$, which shall be redundant. The advantage of doing so is that, unlike the lagrangian without $s$, which shifts by a total derivative proportional to $ B\dot{x}$ under a translation in $y$, the lagrangian in (\ref{Action_LaH}) is genuinely invariant under a central extension by $U(1)$ of the translation group.
 
This central extension is the Heisenberg group, $\mathrm{Hb}$, defined as the equivalence classes of $(x,y,s) \in \mathbb{R}^3$ under the equivalence relation $s \sim s + 2\pi$, with multiplication law
\begin{equation}\label{Hiesenberg_Multiplication}
[(x',y',s')]\cdot [(x,y,s)]=[(x+x',y+y',s+s'-By'x)],
\end{equation}
and corresponding to $\mathbb{R}^2\times S^1$ as a manifold.
Notice that the group $\mathbb{R}^2$ of translations appears not as a {\em sub}group of $\mathrm{Hb}$, but rather as the {\em quotient} group of $\mathrm{Hb}$ with respect to the central $U(1)$ subgroup $\{[(0,0,s)]\}$. Thus we have a homomorphism $\mathrm{Hb} \rightarrow \mathbb{R}^2$, given explicitly by $[(x,y,s)] \mapsto (x,y)$, whose kernel is the central $U(1)$.
Notice that our definition of the group multiplication law depends on $B\in \mathbb{R}$, reflecting the fact that even though the groups with distinct values of $B$ are isomorphic as groups, they are not isomorphic as central extensions.

Given  (\ref{Action_LaH}), the momentum $p_s$ conjugate to $s$ satisfies the constraint $p_s+1=0$. We take care of this in the usual way, by forming the total hamiltonian (see \emph{e.g.}~\cite{henneaux1994quantization})
\begin{equation}
H=\frac{1}{2}\left(p_x+By\right)^2+\frac{1}{2}p_y^2+v(t)\left(p_s+1\right),
\end{equation}
with $p_x$ and $p_y$ being the momenta conjugate to $x$ and $y$ respectively, and with $v(t)$ being a Lagrange multiplier. Upon quantizing (something we will later define formally), we obtain the hamiltonian operator
\begin{equation} \label{Landau_QH}
\hat H=\frac{1}{2} \left(-i\frac{\partial}{\partial x}+B y\right)^2-\frac{1}{2} \frac{\partial^2}{\partial y^2}+v(t)\left(-i\frac{\partial}{\partial s}+1\right),
\end{equation}
which has a natural action on the space of square integrable functions on the Heisenberg group, $L^2(\mathrm{Hb})$. The physical Hilbert space $\mathcal{H}$ must take account of the constraint (or, equivalently, the redundancy in our description), so we define it to be not $L^2(\mathrm{Hb})$, but rather the subspace
\begin{equation} \label{landau hilbert space}
\mathcal{H} = \left\{ \Psi(x,y,s) \in L^2(\mathrm{Hb}) \left| \left(-i\frac{\partial }{\partial s}+1\right)\Psi(x,y,s)=0 \right. \right\}.
\end{equation}
Note that this subspace of $L^2(\mathrm{Hb})$ is closed under the action of the Heisenberg group and under the action of $\hat H$, implying that it is also closed under time evolution.

We then want to solve the time-independent Schr\"{o}dinger equation (from hereon `SE') $\hat H\Psi =E\Psi$. 
To solve the SE, we decompose $\Psi$ into unitary irreducible representations (henceforth `unirreps') of $\mathrm{Hb}$:\footnote{To say we are `decomposing $\Psi$ into unirreps of $\mathrm{Hb}$' is a slight abuse of terminology; what we mean, precisely, is discussed in \S\ref{Sec_QMinMagBack}.} 
\begin{equation} \label{Landa_Decomp}
\Psi(x,y,s)= \int dr dt \frac{|B|}{2\pi}\pi^B(r,t;x,y,s)f(r,t),
\end{equation}
where $r,t\in \mathbb{R}$ are real numbers. Here, 
\begin{equation}\label{Landau_Reps}
\pi^k(r,t;x,y,s)=e^{ik(xr-s/B)} \delta(r+y-t), \qquad k/B \in \mathbb{Z},
\end{equation}
which denote the matrix elements of the infinite-dimensional unirreps of $\mathrm{Hb}$, which act on the vector space $L^2(\mathbb{R},dt)$. The fact that only the unirrep with $k=B$ appears in the decomposition (\ref{Landa_Decomp}) follows from enforcing the constraint in (\ref{landau hilbert space}), as we show in Appendix \ref{Rudiments_of_HA}.

Notice that with this decomposition $\Psi(x,y,s)$ may not be square integrable (as the matrix elements of $\pi^B$ themselves are not). As such, once we have found our `solutions' to the SE with this decomposition we must check that they are square integrable (or more generally the limit of a Weyl sequence). This subtlety will be omitted here due to the familiar form our final solutions will take.

Substituting the decomposition (\ref{Landa_Decomp}) into the SE, and using the constraint to eliminate the Lagrange multiplier, yields
\begin{equation}
\frac{|B|}{2\pi} \int dr dt \left(\frac{1}{2}\left(-i\frac{\partial}{\partial x}+By\right)^2-\frac{1}{2} \frac{\partial^2}{\partial y^2}-E\right) f(r,t)e^{i(Bxr-s)} \delta(r+y-t)=0.
\end{equation}
After some straightforward manipulation, this reduces to
\begin{equation} \label{LandauODE}
 \left(\frac{1}{2}B^2 t^2-\frac{1}{2} \frac{\partial^2}{\partial t^2}-E\right)f(r,t)=0.
\end{equation}
This differential equation, which we recognise as the SE for the simple harmonic oscillator, has the solutions
\begin{equation}
f(r,t)=H_n\left(\sqrt{|B|}t\right) e^{-|B| t^2/2}g(r), \quad E=|B|(n+1/2),
\end{equation}
where $H_n(x)$ are the Hermite polynomials and $g(r)$ is an arbitrary function of $r$. The corresponding eigenfunctions are thus
\begin{equation}
\Psi_n(x,y,s)= \frac{|B|}{2\pi} \int dr dt H_n\left(\sqrt{|B|}t\right) e^{-|B| t^2/2}g(r) e^{i(Bxr-s)} \delta(r+y-t).
\end{equation}
We can of course eliminate our redundant degree of freedom, by setting $s=0$ for example, to obtain corresponding wavefunctions living in $L^2(\mathbb{R}^2)$ (more precisely, the wavefunction is described by a section of a Hermitian line bundle).
In the above expression $g(r)$ accounts for the degeneracy in the Landau levels. On choosing $g(r)=\delta(r-\alpha/B)$ for $\alpha\in \mathbb{R}$ (and setting $s=0$) we arrive at familiar solutions to this system, of the form
\begin{equation}
\Psi_{n,\alpha}(x,y)=e^{i\alpha x} H_n\left(\sqrt{|B|}(y+\alpha/B)\right) e^{-\frac{|B|}{2} (y+\alpha/B)^2}.
\end{equation}

Now let us now recap what we have achieved. Certainly, our result for the spectrum is not new; nor are our observations regarding the momentum generators. Rather, what is new is the observation that we can reformulate the problem via a redundant description, in which a central extension of $G$ by $U(1)$ acts on the configuration space of that redundant description, in a way that allows us to solve for the spectrum using methods of harmonic analysis. While this may seem like overkill, it is important to realise that Landau's original method of solution~\cite{Landau1930} only works for this specific system of a particle on $\mathbb{R}^2$ in a magnetic background, and moreover works only in a particular gauge (the `Landau gauge'). It is not at all clear how such an approach could be generalised to  other target spaces (or gauges).  In contrast, as we shall soon see in \S \ref{Formalism}, using harmonic analysis on a central extension can be generalised to any group $G$ acting on any target space manifold $M$, since it exploits the underlying group-theoretic structure of the system.

\subsection{Bosonic versus fermionic rigid bodies} \label{Sec_RigidBody}

Our second prototypical example illustrates the approach in a case where one cannot form a globally-defined lagrangian without extending the configuration space by a redundant degree of freedom. This prototype
also provides an example where the relation to magnetic fields is not immediately apparent. 

To wit, we consider the quantum mechanics of a rigid body in three space dimensions, whose configuration space is $SO(3)$, with dynamics invariant under the rotation group. 
Evidently, such a rigid body could be either a boson or a fermion (it could, for example, be a composite made up of either an even or odd number of electrons and protons). If it is a fermion, then its wavefunction should acquire a factor of $-1$ when the body undergoes a complete rotation about some axis and we expect, on general physical grounds, that we can represent this effect via a local lagrangian term. To see how it can be done, we first note that the term should be both $SO(3)$ invariant and topological. It is thus reasonable to guess that it can be written in terms of a magnetic field, or more precisely, a connection on some $U(1)$-principal bundle over $SO(3)$.\footnote{For those readers unfamiliar with principal bundles, we note that a technical understanding should not be necessary to follow the discussion in this Section. Nonetheless, since the notion of a principal bundle shall be central to the general formalism which we shall set out in \S \ref{Formalism}, we provide a more-or-less self-contained introduction to the relevant concepts in Appendix \ref{sec_useful_mathematics}.
}  Confirmation that this is indeed the case comes from the fact that (up to equivalence), there are just two $U(1)$-principal bundles over $SO(3)$ (to see this, note that such bundles are classified by the first Chern class, which is a cohomology class in $H^2(SO(3),\mathbb{Z}) \cong \mathbb{Z}/2$). Thus we have the trivial bundle $SO(3)\times U(1)$ and a non-trivial bundle, which we may take to be $U(2)$, the group of $2\times 2$ unitary matrices. Clearly, these are not only $U(1)$-principal bundles, but also they have the structure of central extensions of $SO(3)$ by $U(1)$, which we need for our construction. The trivial bundle admits the zero connection and describes the boson, while the non-trivial bundle admits a non-zero (but nevertheless flat) connection, which accounts for the fermionic phase.

Let us now see this more clearly by means of an explicit construction.
An element $U\in U(2)$ projects down to an element $O\in SO(3)$ by projecting out its ($U(1)$-valued) overall phase. We parameterize a matrix $U\in U(2)$ by
\begin{equation}
U=e^{i\chi}\begin{pmatrix}
e^{i(\psi+\phi)/2}\cos(\theta/2) & e^{-i(\psi-\phi)/2} \sin(\theta/2) \\
-e^{i(\psi-\phi)/2}\sin(\theta/2) & e^{-i(\psi+\phi)/2} \cos(\theta/2)
\end{pmatrix},  
\end{equation}
where $\theta \in[0,\pi]$, $\phi\in[0,2\pi)$, $\psi\in[0,4\pi)$ and $\chi\in [0,2\pi)$ with the equivalence relation $(\theta,\phi,\psi,\chi)\sim(\theta,\phi,\psi+2\pi,\chi+\pi)$.
Now, consider the curve $\gamma^\prime(t)$ in $U(2)$ defined by
\begin{equation}
 \gamma^\prime (t)=\begin{pmatrix} e^{it} & 0 \\ 0 & e^{- it} \end{pmatrix}, \quad t\in[0,\pi],
\end{equation}
and define the curve $\gamma(t)$ to be the projection of $\gamma^\prime(t)$ to $SO(3)$, which one might think of as the particle worldline in the original configuration space.  The curve $\gamma^\prime(t)$ is a horizontal lift of $\gamma(t)$ with respect to the connection, which in our coordinates can be represented by $A=d\chi$. For our purposes here, this simply means that the tangent vector $X_{\gamma^\prime}$ to the curve $\gamma^\prime(t)$ satisfies $A(X_{\gamma^\prime})=0$, \emph{i.e.} it has no component in the $\chi$ direction.

Notice that in $U(2)$ we have $\gamma^\prime(0)=I$ and $\gamma^\prime(\pi)=-I$, and that these two points, while distinct in $U(2)$, both project to the identity in $SO(3)$. The relative phase of $\pi$ between $\gamma^\prime(0)$ and $\gamma^\prime(\pi)$ is called the holonomy of $\gamma(t)$. This implies that the rigid body is in this case a fermion, because the loop $\gamma(t)$ in $SO(3)$ corresponds to a $2\pi$-rotation about the $z$-axis in $\mathbb{R}^3$. If we had instead equipped the rigid body with the trivial choice of bundle $SO(3)\times U(1)$, instead of $U(2)$, then the phase returns to zero upon traversing any closed loop in $SO(3)$, thus corresponding to a boson.

This fermionic versus bosonic nature is furthermore manifest in the differing representation theory of the Lie groups $U(2)$ and $SO(3)\times U(1)$. This shall be important when we solve for the spectrum of this quantum mechanical system in \S \ref{Rigid_Rotation}. While the unirreps of $SO(3)\times U(1)$ are all odd-dimensional (as we would expect for the integral angular momentum eigenstates of a bosonic rigid body), $U(2)$ also contains unirreps of even dimension (for example, the defining 2-d representation), leading to the possibility of eigenstates with half-integral angular momentum, which is exactly what we expect for a fermionic rigid body, via the spin-statistics theorem.

For our purposes, it will be useful to consider a different path $\tilde \gamma(t)$ in $U(2)$ that also projects down to $\gamma$ in $SO(3)$, defined by
\begin{equation}
\tilde \gamma(t)=\begin{pmatrix} e^{ 2it} & 0 \\ 0 & 1 \end{pmatrix}, \quad t\in[0,\pi].
\end{equation} 
While this path $\tilde \gamma$ is not a horizontal lift of the worldline $\gamma$, it nonetheless still projects down to $\gamma$, but is now a closed loop in $U(2)$ with the property that the exponential of the integral over $\tilde \gamma$ of the connection $A=d\chi$ is equal to the holonomy, {\em viz.} $e^{- i \int_{\tilde \gamma} A}=e^{-  i \int_0^\pi dt}=-1$. This means that we can represent the holonomy (which is the contribution to the action phase from the topological term) in terms of a local action, namely the integral of the connection over an appropriately chosen loop $\tilde \gamma$. Given the existence of the horizontal lift, the fact that $U(1)$ is connected means such a loop always exists. As we might expect from the fact that there is a redundancy in our description, the choice of loop is, however, not unique. Nevertheless, the integral is of course independent of this choice.

The upshot is that this topological phase, which results in fermionic statistics of the rigid body, can be obtained from the integral of a lagrangian (the connection) on the principal bundle, here $U(2)$, which is both globally-defined and manifestly local. Due to the topological twisting of the bundle, there is no corresponding globally-defined lagrangian on the original configuration space, here $SO(3)$.

In this Section we have discussed two quantum mechanical prototypes, which are at first sight very different from a physical perspective. What both examples have in common is the possibility of a topological term in the action phase. In our first example of quantum mechanics on the plane (\S \ref{Sec_LandauLevels}), this topological term corresponded to the familiar coupling of our particle to a magnetic field transverse to the plane of motion. We saw that, in order to identify a symmetry group that commutes with the hamiltonian, it was necessary to pass to an equivalent description on an extended space, with that symmetry group being the Heisenberg group. We then saw how one could obtain the Landau level spectrum by using harmonic analysis on the Heisenberg group, a method that works in any gauge. In contrast, in our second example of a rigid body (in this subsection), the topological term corresponded to a vanishing magnetic field, but we nonetheless saw that the term can have interesting effects, in this case leading to either fermionic or bosonic character of the rigid body.

Mathematically, both examples admit a common description: the topological term in the action phase is the holonomy of a connection on a $U(1)$-principal bundle $P$ over the configuration space $M$. Such a topological term may not correspond to any globally-defined lagrangian on $M$ (as in \S \ref{Sec_RigidBody}), or may not be invariant under the action of the group $G$ which acts on $M$ (as in \S \ref{Sec_LandauLevels}); or, indeed, both (interconnected) issues may arise. Having demonstrated in our two prototypes that these problems can be remedied by passing to an equivalent description on an extended space (namely, the principal bundle $P$) with an action by a central extension of $G$, we are now ready to explain the general formalism.

\section{Formalism} \label{Formalism}

We shall consider quantum mechanics of a point particle whose configuration space is a smooth, connected manifold $M$. This can be described by an action whose degrees of freedom are maps $\phi$ from the $1$-dimensional worldline, $\Sigma$, to the target space $M$, \emph{viz.} $\phi: \Sigma\rightarrow M$.  We consider the smooth action $\alpha:G\times M\rightarrow M$ of a connected Lie group $G$ on $M$, which shall define the (global) symmetries of the system. Since, in the path integral approach to quantum mechanics, it is only the \emph{relative} action phase between pairs of worldlines that is physical, we are free are to consider only worldlines which are closed, without loss of generality.
\subsection{Quantum mechanics in magnetic backgrounds} \label{Sec_QMinMagBack}
We will now define the dynamics of the particle on $M$ by specifying a $G$-invariant action phase, $e^{i S[\phi]}$, defined on all closed worldlines, or equivalently on all piecewise-smooth loops in $M$. 

The action consists of two pieces (ignoring potential and higher-derivative terms). The first piece is the kinetic term, constructed out of a $G$-invariant metric on $M$. The second piece in the action couples the (electrically charged) particle to a background magnetic field. This is a topological term in the action phase (in the sense that it does not require the metric), equal to the holonomy of a connection $A$ on a $U(1)$-principal bundle $P$ over $M$ (see Appendix~\ref{sec_useful_mathematics}), evaluated over the loop $\phi$. It is shown in \cite{Davighi2018} that for this term in the action phase to be invariant under the action $\alpha$ of the Lie group $G$, we require that the contraction of each vector field $X$ generating $\alpha$ with the curvature $2$-form $\omega$ is an exact 1-form. That is, we require
\begin{equation}\label{Manton condition}
\iota_{X}\omega=df_X \quad \forall X \in \mathfrak{g},
\end{equation}
where each $f_X$ is a globally-defined function (equivalently, a $0$-form) on $M$. This condition, which we shall refer to as the Manton condition, is necessary for the $G$-invariance of the topological term evaluated on all piecewise-smooth loops in $M$ (provided that $G$ is connected, as we are assuming). This Manton condition is analogous to the moment map formula for a group action to be hamiltonian with respect to a given symplectic structure. The difference here, mathematically, is that the field strength $\omega$ need not be a non-degenerate 2-form.

It will be of use later, when we end up constructing an equivalent action on $P$, to specify a local trivialisation of $P$ over a suitable set of coordinate charts $\{U_\alpha\}$ on $M$. We let $s_\alpha\in[0,2\pi)$ be the $U(1)$-phase in this local trivialisation and define the transition functions $t_{\alpha \beta}=e^{i(s_\alpha -s_\beta)}$.
Technically speaking, we need two coordinate charts  on $P$, denote them $V_{\alpha,1}$ ($s_\alpha \ne \pi$) and $V_{\alpha,2}$ ($s_\alpha \ne 0$), for each $U_\alpha$, to cover the $S^1$ fibre. In what follows, we will often gloss over this technicality; from hereon, $s_\alpha$ should  be assumed to be written locally in one of these coordinate charts, which we shall denote collectively by $V_\alpha$ to avoid drowning in a sea of indices. Following this ethos, we will also tend to drop the $\alpha$ subscript on $s_\alpha$ when we turn to solving the examples in \S \ref{Sec_Examples}.

Our objective is to solve the SE corresponding to this $G$-invariant quantum mechanics, which we shall ultimately achieve by passing to a central extension of $G$ by $U(1)$, and using harmonic analysis on that central extension.

To motivate our method, we shall first review how harmonic analysis can be used to solve the corresponding (time-independent) SE in the {\em absence} of the magnetic background, by exploiting the group-theoretic structure of the system~\cite{Gripaios2016}. Solving the SE amounts to finding the spectrum of an appropriate hamiltonian operator $\hat H$, which in this case can be quantized as the Laplace-Beltrami operator corresponding to the choice of $G$-invariant metric on $M$, on an appropriate Hilbert space. In the absence of a magnetic field, the Hilbert space can be taken to be $L^2(M)$. We can endow this Hilbert space with a highly reducible, unitary representation of $G$, namely the left-regular representation defined by
\begin{equation}
\rho(g)\Psi(m):=\Psi(\alpha_{g^{-1}}m) \text{ for $m\in M$, $g \in G$, and $\Psi\in L^2(M)$}.
\end{equation}
The action of $\rho$ allows us to decompose the vector space $L^2(M)$ into a direct sum (or, more generally, a direct integral) of vector spaces $V^{\lambda,t}$, such that the restriction of $\rho$ to each $V^{\lambda,t}$ yield a unirrep of $G$, which we label by its equivalence class $\lambda\in \Lambda$. Each unirrep may, of course, appear more than once in the decomposition of $L^2(M)$ and so we index these by $t\in T^\lambda$.
We will fix a basis for each vector space $V^{\lambda,t}$, which we denote by $e^{\lambda,t}_r$,
where $r \in R^{\lambda}$ indexes the (possibly infinite-dimensional) basis, which does not depend on $t$. 

In our examples we often specify the operator in the unirrep $\lambda$ by its form in the chosen basis, which we denote $\pi^\lambda(s,q)$, where $s$ and $q$ index the basis. In many cases, as in \S\ref{Sec_LandauLevels}, it will transpire that we 
can set $e^{\lambda,t}_r=\pi^\lambda(r,t)$. In other instances were this is not the case, one can nonetheless infer a suitable form for the $e^{\lambda,t}_r$ from $\pi^\lambda(s,q)$.

It is then a consequence of Schur's lemma that if
\begin{equation} \label{HA_Commutator}
\hat H \rho(g) f(m)=\rho(g) \hat H f(m),
\end{equation}
then the operator $\hat{H}$ will be diagonal in both $\lambda$ and $r$, and can only mix $e^{\lambda,t}_r$ in the index $t$ and not $r$ or $\lambda$, \emph{i.e.} it only mixes between equivalent unirreps. In most cases this simplifies the SE by reducing the number of different types of partial derivatives present, often resulting in a family of ODEs~\cite{Gripaios2016}.

\subsection{An equivalent action with manifest symmetry and locality}

Interestingly, coupling our particle on $M$ to a magnetic background, in the manner described in \S\ref{Sec_QMinMagBack}, may prevent one from constructing a local hamiltonian that satisfies~(\ref{HA_Commutator}). As elucidated by our pair of prototypes in \S \ref{Sec_Prototyes}, there are two obstructions to this method.

Firstly, as demonstrated by our prototypical example (\S\ref{Sec_RigidBody}), it may not be possible to form a globally-valid lagrangian on $M$. Secondly, as demonstrated by our prototypical example (\S\ref{Sec_LandauLevels}), even when the construction of a globally-valid lagrangian is possible (\emph{i.e.} when $\omega$, the magnetic field strength, is the exterior derivative of a globally-defined 1-form), the lagrangian may vary by a total derivative under the action of $G$. This means that (\ref{HA_Commutator}) will fail to hold, and the hamiltonian will not act only between equivalent unirreps of $G$. 

It is possible to overcome both these problems by considering an equivalent dynamics on the principal bundle $\pi:P\rightarrow M$, instead of on $M$, as we shall now explain.

The topological term, which is just the holonomy of the connection $A$ on $P$, can be written as the integral of $A$ over any loop $\tilde \phi$ in $P$ which projects down to our original loop $\phi$ on $M$, \emph{i.e.} one that satisfies $\pi \circ \tilde \phi=\phi$
(see Appendix \ref{sec_useful_mathematics}). Pulling back $A$ to the worldline using $\tilde \phi$, we obtain on a patch $V_{\alpha}$ of $P$
\begin{equation}
\tilde \phi^* A=\left(\dot s_\alpha(t)+A_{\alpha,i}\left(x^k(t)\right)\dot x^i(t)\right)dt,
\end{equation}
where  $x^i(t)\equiv x^i (\pi\circ \tilde \phi(t))$ denote local coordinates in $M$ (with $i=1,\dots,\mathrm{dim}\ M$), $s_\alpha(t)\equiv s_\alpha(\tilde \phi(t))$, $\dot s_\alpha\equiv ds_\alpha / dt$ \&\emph{c}, and $\left.A\right|_{V_{\alpha}}\equiv ds_\alpha+A_{\alpha,i}dx^i$ is the connection restricted to the patch $V_\alpha$. 
Given that we can also pull back the metric, and thus the kinetic term, from $M$ to $P$, we can `lift' our original definition of the action from $M$ to the principal bundle $P$. The contribution to the action from a local patch $V_\alpha$ is then
\begin{equation}\label{ActionP}
\left.S[\tilde \phi]\right|_{V_{\alpha}}=\int dt\;\left\{ g_{ij} \dot x^i \dot x^j-\dot s_\alpha-A_{\alpha,i}\dot x^i\right\} ,
\end{equation}
where $g_{ij}dx^i dx^j$ will henceforth denote the pullback of the metric to $P$.

As we have anticipated, this reformulation of the dynamics on $P$ has two important virtues. Firstly, there is a globally-defined lagrangian $1$-form on $P$ for the topological term, namely the connection $A$. Secondly, this lagrangian is strictly invariant under the Lie group central extension $\tilde G$ of $G$ by $U(1)$, defined to be the set
\begin{equation}
\tilde G=\{ (g,\varphi)\in G\times \mathrm{Aut}(P,A)\mid \pi \circ \varphi=\alpha_g\circ \pi\},
\end{equation}
endowed with the group action $(g,\varphi)\cdot (g^\prime,\varphi^\prime)=(gg^\prime,\varphi\circ \varphi^\prime)$~\cite{neeb2003flux,TUYNMAN1987207}, which as a manifold is the pullback bundle of $\pi:P\rightarrow M$ by the orbit map $\phi_{m}:G\rightarrow M$, $g\mapsto g\cdot m$, for any $m\in M$~\cite{neeb2003flux}. Here, $\mathrm{Aut}(P,A)$ denotes the group of principal bundle automorphisms of $P$ (\emph{i.e.} diffeomorphisms which commute with the right action of the structure group on $P$) which preserve $A$, \emph{i.e.} for $\varphi \in \mathrm{Aut}(P,A)$ we have $\varphi^* A=A$. There is a short exact sequence
\begin{equation}
\begin{tikzcd}
0\arrow{r}{}& U(1)\arrow{r}{\iota}&\tilde G \arrow{r}{\pi^\prime} &G \arrow{r}{} &0,
\end{tikzcd}
\end{equation}
with the subgroup $\mathrm{Im}(\iota)$ central in $\tilde G$, thus exhibiting $\tilde G$ as a central extension of $G$ by $U(1)$. Here $\iota: U(1)\ni e^{i\theta}\mapsto (\mathrm{id},R_{e^{i\theta}})\in\tilde G$, where $R_g\in \mathrm{Aut}(P,A)$ indicates the right action of $U(1)$ on the bundle $P$, and $\pi^\prime: \tilde G \ni (g,\phi)\mapsto g \in G$.
This group has a natural action on the principal bundle $P$, which we denote by $\tilde \alpha: \tilde G\times P \rightarrow P$, defined by $\tilde \alpha_{(g,\varphi)}p=\varphi(p)$, for $p\in P$. 

The price to pay for these two virtues is that we have introduced a redundancy (which locally comes in the form of an extra coordinate $s_\alpha$) into our description. We must account for this redundancy with an appropriate definition of the Hilbert space, to which we turn in the next subsection.

\subsection{Quantization}

Equipped with this reformulation of the dynamics on $P$, and the extended Lie group $\tilde G$, we are now in a position to construct a local hamiltonian operator and solve for its spectrum by decomposing into unirreps of $\tilde G$.

To do this, we first form the classical hamiltonian by taking the Legendre transform of the lagrangian, defined on the `extended phase space' $T^*P$. At this stage the redundancy in our description becomes apparent, with the momentum $p_{s_\alpha}$ conjugate to the (local) fibre coordinate $s_\alpha$ being constant, \emph{viz.} $p_{s_\alpha}+1=0$, as we saw in \S \ref{Sec_LandauLevels}. We can enforce this constraint by quantizing the so-called `total hamiltonian' 
\begin{equation}
\left.H\right|_{V_{\alpha}}=\frac{1}{2}(p_i +A_{\alpha,i})g^{ij}(p_j+A_{\alpha,j})+v(t)(p_{s_\alpha}+1),
\end{equation}
where $p_i$ is the momentum conjugate to the coordinate $x^i$, and $v(t)$ is an arbitrary function of $t$ which plays the role of a Lagrange multiplier. This hamiltonian is naturally quantized as the magnetic analogue of the Laplace-Beltrami operator, in which the covariant derivative $\nabla$ on $M$ is replaced by $\nabla+A$, giving
\begin{equation}\label{QMHamiltonian}
\left.\hat H\right|_{V_{\alpha}}=\frac{1}{2}\left(-i\frac{1}{\sqrt{g}}\frac{\partial}{\partial x^i} \sqrt{g} +A_{\alpha,i}\right)g^{ij}\left(-i\frac{\partial}{\partial x^j}+A_{\alpha,j}\right)+v(t)\left(-i\frac{\partial}{\partial s_{\alpha}}+1\right),
\end{equation}
which is a Hermitian operator acting on the Hilbert space
\begin{equation}
\mathcal{H}=\left\{\Psi\in L^2(P,\tilde \mu)\left| \left(-i\frac{\partial}{\partial s_{\alpha}}+1\right)\Psi=0 \text{ on $V_{\alpha}$} \right.\right\}
\end{equation}
where locally the measure is given by $\tilde \mu=\sqrt{g}\ dsdx^1\ldots dx^n$. The Hilbert space $\mathcal{H}$ is isomorphic to the space of square integrable sections on the hermitian line bundle associated with $P$ with respect to the measure $\mu=\sqrt{g}\ dx^1\ldots dx^n$~\cite{TUYNMAN1998607,WU1976365}.

\subsection{Method of solution: harmonic analysis on central extensions} \label{Method_of_Solution}

Because the local hamiltonian commutes with the left regular representation of $\tilde G$,
we expect to be able to use harmonic analysis on $\tilde G$ (when it exists!) to solve for the spectrum of (\ref{QMHamiltonian}).
The Hilbert space $\mathcal{H}$ is endowed with the left-regular representation $\rho$ of $\tilde G$, under which a wavefunction $\Psi\in\mathcal{H}$ transforms as
\begin{equation}\label{LRR}
\tilde \rho(\tilde g)\Psi(p)\equiv \Psi(\tilde \alpha_{\tilde g^{-1}}p)\qquad \forall p\in P,\ \tilde g \in \tilde G.
\end{equation}
We use harmonic analysis to decompose this representation into unirreps of $\tilde G$, in analogy with how we decomposed into unirreps of $G$ in the absence of a magnetic background, above. Thus, let $e^{\lambda,t}_r(p\in P)$ now denote a basis for this decomposition, which schematically takes the form
\begin{equation} \label{general decomp}
\Psi = \sum_\lambda \int \mu(\lambda,r,t) f^\lambda(r,t) e^{\lambda,t}_r(p) \in L^2(P,\tilde \mu) 
\end{equation}
for an appropriate measure $\mu(\lambda,r,t)$. Note that the basis functions may not be square integrable; if this is not the case one may check that the solutions are the limit of an appropriate Weyl sequence (see \emph{e.g.} \cite{Gripaios2016}). In the presence of the magnetic background, we have passed to a redundant formulation of the dynamics on $P$, and the crucial difference is that we must now account for this redundancy when using harmonic analysis. It turns out (see Appendix \ref{Rudiments_of_HA}) that this redundancy can often be accounted for by restricting the decomposition in (\ref{general decomp}) to the subspace of unirreps which satisfy the constraint $(-i\partial_s +1)e^{\lambda,t}_r(p)=0$, which we can moreover equip with an appropriate completeness relation. In the examples that follow in \S \ref{Sec_Examples}, this decomposition into a restricted subspace of unirreps will serve as our starting point for harmonic analysis.

Then, exactly as above, the fact that the hamiltonian commutes with the left-regular representation (of $\tilde G$, not $G$) means that the action of $\hat{H}$ will only mix equivalent representations (that is, it can mix between different values of the $t$ index, but not the $r$ index or $\lambda$ label). Thus, the SE will be simplified, often to a family of ODEs, as we shall see explicitly in a plethora of examples in the following Section.

It is important to acknowledge that performing harmonic analysis in the manner we have described, for the general setup of interest in which a (possibly non-compact) general Lie group acts non-transitively on the underlying manifold, is far from being a solved problem in mathematics. For example, it is not known under what conditions the integrals denoted in (\ref{general decomp}) actually exist, and whether the functions $f^{\lambda}(r,t)$ can be extracted from $\Psi$ by appropriate integral transform methods. Thus, much of what has been said should be taken with a degree of caution. Fortunately, in the examples that we consider in \S \ref{Sec_Examples}, all of the required properties follow from properties of the usual Fourier transform, and in all cases the method that we have outlined in this section works satisfactorily.

\section{Examples} \label{Sec_Examples}
In \S\S \ref{Sec_LandauLevels} and \ref{Sec_RigidBody} we explained the use of our method for planar motion in a magnetic field, then pointed out the existence of a topological term for the quantum mechanical rigid body, and explained how this term can endow the rigid body with fermionic statistics. We will start this Section where \S \ref{Sec_RigidBody} left off, by solving for the spectrum of this fermionic rigid body using harmonic analysis on the group $U(2)$. After this we will look at a series of other examples where our method is of use. Some of these are well known systems, \emph{e.g.} charged particle motion in the field of a Dirac monopole, whilst others are new, \emph{e.g.} the motion of a particle on the Heisenberg manifold. The results of all the examples considered in this paper are summarised in Table \ref{Example_table}.

\renewcommand{\arraystretch}{1.5} 

\begin{table} 

\begin{center}
 \begin{tabular}{|>{\footnotesize}p{0.52in}|>{\footnotesize \centering}p{0.5in}|>{\footnotesize \centering}p{0.85in}|>{\footnotesize}p{2.40in}|>{\footnotesize}p{1in}|} 
\hline
\S &\thead{$M$\\ $[G]$}& \thead{$P$\\ $[\tilde G]$} &Lagrangian on $P$ & Spectrum\\ \hline
\ref{Sec_LandauLevels} \newline Landau \newline levels & \thead{$\mathbb{R}^2$\\  $[\mathbb{R}^2]$} & \thead{$\mathbb{R}^2\times U(1)$ \\ $[\mathrm{Hb}]$} & $\frac{1}{2} \dot x^2+\frac{1}{2} \dot y^2- \dot s- B y \dot x$ &$|B|(n+1/2)$, \newline $n\in \mathbb{N}_0$\\ \hline
\ref{Rigid_Rotation} \newline Fermionic \newline rigid \newline body & \thead{$\mathbb{R}P^3$\\ $[SO(3)]$}& \thead{$U(2)$\\ $[U(2)]$} & $\frac{1}{2}\left(\dot \theta^2+\dot \phi^2 \sin^2(\theta)+ \left(\dot \psi+\dot \phi \cos(\theta)\right)^2\right)-\dot s$ &$j(j+1)/2$,\newline $j\in \mathbb{N}_0+1/2$\\ \hline
\ref{Sec_DiracMonopole} \newline Dirac \newline monopole & \thead{$S^2$ \\$[SU(2)]$} &\thead{ $L(g,1)$ \\ $[SU(2)\times U(1)]$ } & $\frac{1}{2}\left( \dot \theta^2+\sin^2(\theta) \dot \phi^2\right)-\frac{1}{2}\dot{\chi}-\frac{g}{2} \cos(\theta) \dot \phi$ &$\frac{1}{8}(4j^2+4j-g^2)$,\newline $j \in \mathbb{N}_0+g/2$ \\ \hline
\ref{Sec_dyon} \newline Dyon  & \thead{$\mathbb{R}_+\times S^2$\\$[SU(2)]$} & \thead{$\mathbb{R}_+\times L(g,1)$\\$[SU(2)\times U(1)]$} & $\frac{1}{2}\left( \dot \theta^2+\sin^2(\theta) \dot \phi^2\right)-\frac{q}{r}-\frac{1}{2}\dot{\chi}-\frac{g}{2} \cos(\theta) \dot \phi$ &$-q^2/(2(n+a))$,\newline$n\in \mathbb{N}_{>0}$,\newline $a=\frac{1}{2}(1+((2j+1)^2-g^2)^{1/2})$\\ \hline
\ref{Sec_LandauISO} \newline Landau \newline levels &\thead{$\mathbb{R}^2$\\$[ISO(2)]$}&\thead{ $\mathbb{R}^2\times U(1)$\\ $[\widetilde{\mathrm{ISO}}(2)]$} &$\frac{1}{2}(\dot x^2+\dot y^2)-\dot s - \partial_x h(x,y)\dot x-\partial_y h(x,y) \dot y-B y \dot x$ &$|B|(n+1/2)$, \newline $n\in \mathbb{N}_0$  \\ \hline
\ref{Sec_Heisenberg} &\thead{$\mathbb{R}^3$\\ $[\mathrm{Hb}]$} & \thead{$\mathbb{R}^4$ \\ $[\widetilde{\mathrm{Hb}}]$ } &$ \frac{1}{2}( \dot x^2+ \dot y^2+(\dot z-x \dot y)^2)-\dot s -x \dot z+ \frac{x^2}{2} \dot y$ & Anharmonic oscillator \\ \hline
\ref{Sec_varymass} & \thead{$\mathbb{R}^3$\\ $[\mathbb{R}^2]$}&\thead{$\mathbb{R}^3\times U(1)$\\$[\mathrm{Hb}]$} &$\frac{1}{2}\left(\frac{1}{a+z^2}\dot x^2+\frac{1}{a+z^2} \dot y^2+ \dot z^2\right)-\dot s-B y \dot x$ &$\sqrt{|B|(2n+1)}(m+1/2)+a|B|(n+1/2)$, \newline $n,m \in \mathbb{N}_0$\\ \hline
\end{tabular}
\caption{Summary of examples presented in this paper. The particle lives on the manifold $M$, with dynamics invariant under $G$. Coupling to a magnetic background defines a  $U(1)$-principal bundle $\pi:P\rightarrow M$, on which we form a lagrangian strictly invariant under a $U(1)$-central extension of $G$, denoted $\tilde G$.}
\label{Example_table}
\end{center}
\end{table}
\subsection{Back to the rigid body} \label{Rigid_Rotation}

We resume the example discussed in \S \ref{Sec_RigidBody}. On a local coordinate patch on $P=U(2)$, we define a $U(2)$-invariant action incorporating a kinetic term by
\begin{equation} \label{Rigid_body3}
S=\int dt \left( \frac{1}{2}\dot \theta^2+\frac{1}{2}\dot \phi^2 \sin^2\theta+\frac{1}{2} \left(\dot \psi+\dot \phi \cos\theta\right)^2-\dot s \right).
\end{equation}
The total hamiltonian on this patch is
\begin{equation}
H=\frac{1}{2} p_\theta^2+\frac{1}{2\sin^2\theta}\left( p_\phi^2+p_\psi^2-2\cos\theta\ p_\phi p_\psi\right)+v(t)(p_s+1),
\end{equation}
which we quantize as the operator
\begin{multline} \label{Rigid_BodyQH}
\hat H=-\frac{1}{2\sin\theta} \frac{\partial}{\partial \theta} \left(\sin\theta \frac{\partial}{\partial \theta}\right)-\frac{1}{2\sin^2\theta}\left(\frac{\partial^2}{\partial \psi^2}+\frac{\partial^2}{\partial \phi^2}-2\cos\theta \frac{\partial^2}{\partial \phi\partial \psi}\right)\\+v(t)\left(-i\frac{\partial}{\partial s}+1\right),
\end{multline}
acting on wavefunctions $\Psi(\theta,\phi,\psi,s)\in L^2 (U(2))$ satisfying $\left(-i\frac{\partial}{\partial s}+1\right)\Psi = 0$.
The unirreps whose matrix elements satisfy this condition when considered as functions on $U(2)$, are given by
\begin{equation}
\pi^j_{m,m^\prime}(\theta,\phi,\psi,s)=e^{-is} D^j_{m^\prime m}(\theta,\phi,\psi),
\end{equation}
where $j$ is a positive half-integer, $m$,  $m^\prime \in \{-j,-j+1,\ldots,j\}$, and $D^j_{m^\prime m}$ is a Wigner D-matrix, defined (in our local coordinates) by
\begin{multline}\label{Wigner_D} D^j_{m^\prime m}(\theta,\phi,\psi)=\left( \frac{(j+m)!(j-m)!}{(j+m^\prime)!(j-m^\prime)!}\right)^{1/2} (\sin(\theta/2))^{m-m^\prime}(\cos(\theta/2))^{m+m^\prime} \\ P^{(m-m^\prime,m+m^\prime)}_{j-m}(\cos\theta) e^{-i m^\prime \psi} e^{-i m \phi}.\end{multline}
These are matrix elements of an unirrep of $U(2)$ and, as was the case in \S\ref{Sec_LandauLevels}, transform in the corresponding conjugate representation when the left-regular representation is applied.
The Wigner D-matrices satisfy the completeness relation
\begin{multline}
\sum_{m^\prime\in \mathbb{Z}+1/2}\sum_{m\in \mathbb{Z}+1/2}\sum_{j=\mathrm{max}(|m|,|m^\prime|)}^{\infty}\frac{2j+1}{8\pi^2} \left(D^j_{m^\prime m}(\theta^\prime,\phi^\prime,\psi^\prime)\right)^*D^j_{m^\prime m}(\theta,\phi,\psi)\\= \delta_{2\pi}(\phi-\phi^\prime)\delta_{2\pi}(\psi-\psi^\prime)\delta(\cos\theta-\cos\theta^\prime),
\end{multline}
where $\delta_{2\pi}(\cdots)$ represents a Dirac delta comb with periodicity $2\pi$,  and the sum over $j$ is over half-integers.

Following the formalism set out in \S \ref{Formalism}, we decompose $\Psi$ into a basis $\{e^{j,m^\prime}_m\}$ for $L^2 (U(2))$, which in this case can be chosen to be $e^{j,m^\prime}_m=\pi^j_{m,m^\prime}$, the matrix elements of unirreps of $U(2)$ introduced above, giving us
\begin{equation} \label{Rigid_Bodydecom}
\Psi=\sum_{m^\prime\in \mathbb{Z}+1/2}\sum_{m\in \mathbb{Z}+1/2}\sum_{j=\mathrm{max}(|m|,|m^\prime|)}^{\infty}\frac{2j+1}{8\pi}e^{-is} D^j_{m^\prime m}(\theta,\phi,\psi)f^j_{m^\prime m},
\end{equation}
with inverse
\begin{equation}
f^j_{m^\prime m}=\int d\left(\cos(\theta^\prime)\right) d\psi^\prime d\phi^\prime \left(D^j_{m^\prime m}(\theta^\prime,\phi^\prime,\psi^\prime) e^{-is} \right)^* \Psi(\theta^\prime,\phi^\prime,\psi^\prime,s).
\end{equation}
The SE then reduces to
\begin{equation}
\sum_{m^\prime\in \mathbb{Z}+1/2}\sum_{m\in \mathbb{Z}+1/2}\sum_{j=\mathrm{max}(|m|,|m^\prime|)}^{\infty}\frac{2j+1}{8\pi}\left\{\frac{j(j+1)}{2} -E\right\}e^{-is} D^j_{m^\prime m}(\theta,\phi,\psi)f^j_{m^\prime m}=0,
\end{equation}
yielding the energy levels
\begin{equation}
E^j_{m^\prime m}=\frac{1}{2}j(j+1), \qquad \text{for $j$ half-integer}.
\end{equation}
The corresponding wavefunctions, on our local coordinate patch, can be written 
 \begin{equation}
 \Psi^j_{m^\prime m}(\theta,\phi,\psi,s)=e^{-is}D^j_{m^\prime m}(\theta,\phi,\psi).
 \end{equation}
Setting the fibre coordinate $s$ to zero defines, a section on the hermitian line bundle associated with the principal bundle $U(2)$, in other words a physical wavefunction. On traversing a double intersection of coordinate charts on $SO(3)$, the above expression for the section will shift by a transition function.

We note in passing that on setting $s=0$ the $U(2)$ representations appearing in this decomposition reduce to representations of $SU(2)$. This occurs due to a well-known happy accident, namely that the projective representations of a Lie group $G$ (here $SO(3)$) whose second Lie algebra cohomology vanishes (as is the case for every semi-simple Lie group) in fact correspond to {\em bona fide} representations of the universal cover of $G$ (here $SU(2)$). That is, under these conditions, familiar to most physicists, we may decompose the Hilbert space into unirreps of the universal cover of $G$, without technically needing to pass to a central extension. It is, however, important to point out that even in an example such as this, one cannot write down a local action for the topological term on the universal cover $SU(2)$, but must pass to the central extension, $U(2)$.

\subsection{The Dirac monopole} \label{Sec_DiracMonopole}

Here we consider the $G=SU(2)$-invariant dynamics of a particle moving on the $2$-sphere. We may embed $M=S^2$ in $\mathbb{R}^3$, parametrized by the standard spherical coordinates $(\theta\sim  \theta+\pi,\phi\sim \phi+2\pi)$. We cover $S^2$ with two charts $U_+$ and $U_-$, which exclude the South and North poles respectively. At the centre sits a magnetic monopole of charge $g\in \mathbb{Z}$.
This background magnetic field specifies a particular $U(1)$-principal bundle $P_g$ over $S^2$ with connection $A$, which we may write in our coordinates as
\begin{equation}
\begin{aligned}
\left.A\right|_{U_+}&=ds_+-\frac{g}{2}\left(1-\cos\theta\right)d\phi\\
\left.A\right|_{U_-}&=ds_--\frac{g}{2}\left(-1-\cos\theta\right)d\phi,
\end{aligned}
\end{equation}
where $s_{\pm}$ denotes a local coordinate in the $U(1)$ fibre. This can be conveniently written as
\begin{equation}
A=\frac{1}{2}d\chi+\frac{g}{2}\cos\theta d\phi,
\end{equation}
where $\frac{1}{2}\chi=s_+-\frac{g}{2} \phi$ on $U_+$ and $\frac{1}{2}\chi=s_-+\frac{g}{2} \phi$ on $U_-$.
The transition functions over a trivialisation on $\{U_+,U_-\}$ are specified via the choice
\begin{equation}
(p,e^{i\delta})\in U_+ \times U(1) \mapsto (p,e^{i\delta} e^{ig\phi}) \in U_-\times U(1).
\end{equation}
For general $g$, this bundle $P_g$ is in fact the lens space $L(g,1)$, which is a particular quotient of $S^3$ by a $\mathbb{Z}/g\mathbb{Z}$ action. When $g=1$, the bundle is simply $P_1\cong S^3$, described via the Hopf fibration and when $g=2$, the bundle is simply $\mathbb{R}P^3$.\footnote{The lens spaces $L(g,1)$ make another appearance in physics as the possible vacuum manifolds for the electroweak interaction \cite{Gripaios:2016ubw}.}

As was the case in the previous example, it is here not possible to write down a global $1$-form lagrangian on $S^2$. Rather, as was first demonstrated by Wu \& Yang \cite{wu1976dirac},
one must write the action on $S^2$ as a sum of line integrals on different charts, together with the insertion of $0$-forms (the transition functions) evaluated at points in double intersections of charts.
Thus, it is not possible to use the usual hamiltonian formalism to solve for the spectrum of the corresponding quantum mechanics problem.

Following our formalism, we should instead reformulate the problem by writing down an equivalent, globally-defined lagrangian on the $U(1)$-principal bundle $P_g=L(g,1)$ defined above. The action is
\begin{equation} \label{Action_DiracMonopole}
S=\int dt \left\{ \frac{1}{2}\left( \dot \theta^2+\sin^2\theta\ \dot \phi^2\right)-\frac{1}{2}\dot{\chi}-\frac{g}{2} \cos\theta\ \dot \phi\right\}.
\end{equation}
This lagrangian is invariant under $\tilde G=SU(2)\times U(1)$, the unique (up to Lie group isomorphisms) $U(1)$-central extension of $SU(2)$, with uniqueness following from the fact that $SU(2)$ is a simple and simply-connected Lie group~\cite{TUYNMAN1987207}. We parametrize an element $\tilde g \in \tilde G$ by
\begin{equation}
\tilde g = \left( \begin{pmatrix} e^{i(\psi+\phi)/2} \cos\frac{\theta}{2} & e^{-i(\psi-\phi)/2} \sin\frac{\theta}{2}  \\ - e^{i(\psi-\phi)/2} \sin\frac{\theta}{2}  & e^{-i(\psi+\phi)/2} \cos\frac{\theta}{2}  \end{pmatrix},\ e^{i(g\psi-\chi)/2}\right)\in SU(2)\times U(1).
\end{equation}
The corresponding total hamiltonian is
\begin{equation}
\hat H=\frac{1}{2} p_\theta^2+\frac{1}{2\sin^2\theta}\left(p_\phi+\frac{g}{2} \cos\theta\right)^2+v(t)\left(p_{\chi}+\frac{1}{2}\right),
\end{equation}
which when quantized gives
\begin{equation}
\hat H=-\frac{1}{2\sin\theta} \frac{\partial }{\partial \theta} \left( \sin\theta \frac{\partial}{\partial \theta} \right)+\frac{1}{2\sin^2\theta}\left(-i \frac{\partial}{\partial \phi}+\frac{g}{2} \cos\theta\right)^2+v(t) \left(-i \frac{\partial}{\partial \chi}+\frac{1}{2}\right), \label{monopole ham}
\end{equation}
where the Hilbert space $\mathcal{H}$ is the subspace of square integrable functions on $L(g,1)$ for which the last term in (\ref{monopole ham}) vanishes.

We now wish to solve for the spectrum of this hamiltonian using harmonic analysis on the Lie group $\tilde G = SU(2) \times U(1)$. Matrix elements of unirreps of $SU(2) \times U(1)$ which are annihilated by the constraint $\left(-i \frac{\partial}{\partial \chi}+\frac{1}{2}\right)\pi^j_{m,m^\prime}=0$ are given by
\begin{equation}
\pi^j_{m,m^\prime }(\theta,\phi,\psi,\chi)=e^{i(g\psi-\chi) /2} D^j_{m^\prime m}(\theta,\phi,\psi).
\end{equation}
Here $D^j_{m^\prime m}\equiv e^{-im^\prime \psi-im\phi} d^j_{m^\prime m}(\theta)$ are the same Wigner $D$-matrices as defined in (\ref{Wigner_D}), and the matrices $d^j_{m^\prime m}(\theta)$ are conventionally referred to as  `Wigner $d$-matrices'.
The subspace of these unirreps with $m^\prime=g/2$ do not depend on the coordinate $\psi$, and provide a suitable basis for decomposing square-integrable functions on the lens space $L(g,1)$. We denote these basis functions by $e^{j,g/2}_m(\theta,\phi,\chi)=\pi^{j}_{m,g/2}(\theta,\phi,\psi,\chi)$, which satisfy the constraint condition and which transform as unirreps of $SU(2)\times U(1)$. This subspace of $\mathcal{H}$ carries the completeness relation
\begin{multline}
\sum_{m+g/2\in \mathbb{Z}} \;\sum_{j=\max(|m|,g/2)}^\infty \frac{2j+1}{4\pi} \left(e^{j,g/2}_m(\theta^\prime,\phi^\prime,\chi^\prime)\right)^*e^{j,g/2}_m(\theta,\phi,\chi)\\=e^{-i(\chi-\chi^\prime)/2}\delta_{2\pi}(\phi-\phi^\prime)\delta(\cos\theta-\cos\theta^\prime),
\end{multline}
 which allows us to decompose any wavefunction in $\Psi \in \mathcal{H}$ into unirreps as follows
\begin{equation} \label{Dirac_Decomp}
\Psi(\theta,\phi,\chi)=e^{-i\chi/2}\sum_{m+g/2\in \mathbb{Z}} \;\sum_{j=\max(|m|,g/2)}^\infty \frac{2j+1}{4\pi} f_m^j e^{-im \phi}d^j_{g/2,m} (\theta),
\end{equation}
where
\begin{equation}
f_m^j=\int d(\cos\theta^\prime) d\phi^\prime\; e^{im \phi^\prime +i\chi^\prime/2} d^j_{g/2,m}(\theta^\prime)\Psi(\theta^\prime,\phi^\prime,\chi^\prime).
\end{equation}

If we now substitute the decomposition (\ref{Dirac_Decomp}) into the SE, after simplification, we get
\begin{equation} \label{Dirac_SE}
\sum_{m+g/2\in \mathbb{Z}} \;\sum_{j=\max(|m|,g/2)}^\infty \frac{2j+1}{4\pi} \left(\frac{1}{8}(4j^2+4j-g^2)-E\right)e^{-i\chi/2}e^{-im \phi}d^j_{g/2,m}(\theta)=0.
\end{equation}
Thus the solution to the SE is
\begin{equation}
\Psi^j_m(\theta,\phi, \chi)=e^{-i\chi/2-im \phi}d_{g/2,m}^j(\theta),\quad E^j_m=\frac{1}{8}(4j^2+4j-g^2).
\end{equation}
Notice that the eigenstates are labeled by two quantum numbers $j$ and $m$, but that for a given $j$ the eigenstates with different values of $m$ are degenerate in energy due to the rotational invariance of the problem.

To write our solution in terms of a section on a hermitian line bundle associated with $P_g$, we set $s_+=0$ on $U_+$ and $s_-=0$ on $U_-$, corresponding to $\chi=-g\phi$ and $\chi=g\phi$ respectively. This yields
\begin{equation}
\begin{aligned}
\Psi^j_{m,+}(\theta,\phi)&=e^{i\frac{g}{2} \phi-im \phi}d_{g/2,m}^j(\theta),\\
\Psi^j_{m,-}(\theta,\phi)&=e^{-i\frac{g}{2}\phi-im \phi}d_{g/2,m}^j(\theta).
\end{aligned}
\end{equation}
These solutions agree with the solutions of Wu and Yang \cite{WU1976365}, who solved this system by considering  local hamiltonians on $U_+$ and $U_-$ separately. 

\subsection{Charged particle orbiting a dyon} \label{Sec_dyon}

In the previous Section we found the spectrum of an electrically charged particle in the presence of a magnetic monopole. Within our formalism, it is straightforward to generalize this to study an electrically charged particle in the background field of a dyon, and use harmonic analysis to reduce the corresponding SE to an ODE.

The required modification is to include an $r$-dependent kinetic term, where $r$ is the radial distance from a dyon located at the origin, together with an $r$-dependent potential term, in the action (\ref{Action_DiracMonopole}). We have
\begin{equation} \label{Action_Dyon}
S=\int dt \left\{ \frac{1}{2}\left(\dot r^2+ r^2\dot \theta^2+r^2\sin^2\theta\ \dot \phi^2\right)-\frac{q}{r}-\frac{1}{2}\dot{\chi}-\frac{g}{2} \cos\theta\ \dot \phi\right\}.
\end{equation}
where $q$ is the electric charge of the dyon, and $g\in\mathbb{Z}$ is the (quantized) magnetic charge of the dyon as before. The original configuration space $M$ of the system is $\mathbb{R}_+\times S^2$, whilst this action is written on the $U(1)$-principal bundle $P_{q,g}=\mathbb{R}_+\times L(g,1)$ where $L(g,1)$ is the lens space as in \S \ref{Sec_DiracMonopole}.  This action is invariant under a \emph{non-transitive} action of $SU(2)\times U(1)$, as defined in the previous Section.

The quantized total hamiltonian corresponding to (\ref{Action_Dyon}) is given by
\begin{multline}
\hat H=-\frac{1}{2 r^2} \frac{\partial}{\partial r} \left( r^2 \frac{\partial }{\partial r}\right)-\frac{1}{2\sin\theta} \frac{\partial}{\partial \theta} \left( \sin\theta \frac{\partial}{\partial \theta}\right)-\frac{1}{2 r^2\sin^2\theta} \left(-i\frac{\partial}{\partial \phi} +\frac{g}{2}\cos\theta\right)^2\\
+\frac{q}{r} +v(t) \left(-i \frac{\partial}{\partial \chi}+\frac{1}{2}\right)
\end{multline}
which acts on the physical Hilbert space. The decomposition of a wavefunction $\Psi(r,\theta,\phi,\chi)$ in this Hilbert space is completely analogous to the decomposition in (\ref{Dirac_Decomp}), however this time the $f_m^j$, which where previously constants, should be replaced with functions $f_m^j(r)$. On substituting this decomposition into the SE, we arrive at the following differential equation for $f_m^j(r)$,
\begin{equation}
\left(-\frac{1}{2 r^2} \frac{\partial}{\partial r} \left( r^2 \frac{\partial }{\partial r}\right)+\frac{1}{8r^2}(4j^2+4j-g^2)+ \frac{q}{r} -E\right)f_m^j(r)=0.
\end{equation}
The bounded solutions to this ODE were derived in~\cite{Bose_1985}, giving the spectrum
\begin{equation}
E_n=-\frac{q^2}{2(n+a)^2}, \quad n\in \mathbb{N}_{>0},
\end{equation}
where $a=\frac{1}{2}\left(1+\left((2j+1)^2-g^2\right)^{1/2}\right)$.

\subsection{Planar motion in a uniform magnetic field (take two)} \label{Sec_LandauISO}
In \S \ref{Sec_LandauLevels} we solved for the spectrum of a particle on $\mathbb{R}^2$ in the presence of a uniform magnetic field perpendicular to the plane, by considering the group $\mathbb{R}^2$ of translations in the plane, and passing to its central extension, the Heisenberg group $\mathrm{Hb}$. Of course, the symmetry group of this system is larger than $\mathbb{R}^2$, because both the kinetic term and the magnetic coupling are invariant not just under translations, but also under rotations. Thus, in this Section, we revisit this problem (and solve it again) using a different implementation of our general method, by instead
considering the particle as living on the quotient space $M=\mathrm{ISO}(2)/SO(2)\cong \mathbb{R}^2$, with $G=\mathrm{ISO}(2)$ being the Euclidean group in two dimensions. Thus, our solution here shall involve the representation theory of a central extension of $G=\mathrm{ISO}(2)$, which will be a four-dimensional group, rather than the representation theory of $\mathrm{Hb}$ which was used in \S \ref{Sec_LandauLevels}.

As usual, we formulate the action on a $U(1)$-principal bundle $P$ over the target space $M=\mathrm{ISO}(2)/SO(2)\cong \mathbb{R}^2$. Using coordinates $(x,y,s)$, where $(x,y)\in \mathbb{R}^2$ provide global coordinates on the base space, and $s$ denotes a local coordinate in the $U(1)$ fibre, the action is
\begin{equation}
S=\int\left( \frac{1}{2}(\dot x^2+\dot y^2)-\dot s-\frac{\partial h}{\partial x} \dot x-\frac{\partial h}{\partial y} \dot y -B y \dot x\right)dt,
\end{equation}
where $h(x,y)$ is an arbitrary smooth function of $x$ and $y$, which corresponds to a choice of gauge for the magnetic vector potential. Note that in all the examples in this paper, there is a choice of gauge made in writing down the magnetic vector potential which appears in the action. While different choices of gauge will in general result in different central extensions $\tilde G$, gauge-equivalent vector potentials nonetheless correspond to central extensions which are isomorphic as Lie groups. In this sense, the choice of gauge has little affect on the representation theory used in our calculations. For this example, we have chosen to make this gauge-dependence (or, rather, independence) explicit, by formulating the action in a general gauge from the outset.

As usual, the lagrangian is not invariant under the isometry group $G=\mathrm{ISO}(2)$, but rather it shifts by a total derivative under the translation subgroup. The lagrangian is, however, genuinely invariant under a $U(1)$-central extension of $\mathrm{ISO}(2)$, which we will denote by $\widetilde{\mathrm{ISO}}(2)$, which is a four-dimensional group defined by 
\begin{multline}
\Big\{\xi_x^\prime,\xi_y^\prime,\xi_c^\prime,\xi_s^\prime\Big\}\cdot\Big\{\xi_x,\xi_y,\xi_c,\xi_s\Big\}=\Big\{\xi_x^\prime+\xi_x\cos\xi_c^\prime+\xi_y\sin\xi_c^\prime ,\ \xi_y^\prime+\xi_y\cos\xi_c^\prime-\xi_x\sin\xi_c^\prime,\ \xi_c+\xi_c^\prime,\\\xi_s+\xi_s^\prime-\frac{B}{2}\left((\xi_x\cos\xi_c^\prime+\xi_y\sin\xi_c^\prime)\xi_y^\prime-(\xi_y\cos\xi_c^\prime-\xi_x\sin\xi_c^\prime)\xi_x^\prime\right) \Big\}.
\end{multline}
This group acts on the principal bundle $P$ via
\begin{multline}
\tilde \alpha_{(\xi_x^\prime,\xi_y^\prime,\xi_c^\prime,\xi_s^\prime)}\cdot(x,y,s)=\Big\{x^\prime , y^\prime,\\\xi_s+\xi_s^\prime-\frac{B}{2}\left((x\cos\xi_c^\prime +y\sin\xi_c^\prime)\xi_y^\prime-(y\cos\xi_c^\prime-x\sin\xi_c^\prime)\xi_x^\prime\right)+\left(\frac{B}{2} xy-\frac{B}{2}x^\prime y^\prime\right)+(h(x,y)-h(x^\prime,y^\prime)) \Big\},
\end{multline}
where $x^\prime=\xi_x^\prime+x\cos\xi_c^\prime+y\sin\xi_c^\prime$ and $y^\prime=\xi_y^\prime+y\cos\xi_c^\prime-x\sin\xi_c^\prime$.

The corresponding total hamiltonian is
\begin{equation}
H=\frac{1}{2}\left(p_x+\frac{\partial h}{\partial x} +By\right)^2+\frac{1}{2}\left(p_y+\frac{\partial h}{\partial y} \right)^2+v(t)(p_s+1),
\end{equation}
which we quantize as the Hermitian operator
\begin{equation}
\hat H=\frac{1}{2}\left(-i \frac{\partial}{\partial x}+\frac{\partial h}{\partial x} +By\right)^2+\frac{1}{2}\left(-i\frac{\partial}{\partial y}+\frac{\partial h}{\partial y} \right)^2+v(t)\left(-i\frac{\partial}{\partial s}+1\right).
\end{equation}
The Hilbert space $\mathcal{H}$ is the subspace of square integrable functions on the bundle $P$ which are annihilated by the constraint  $\left(-i\frac{\partial}{\partial s}+1\right)=0$. We shall now solve the SE for this system by decomposing this Hilbert space into unirreps of the group $\widetilde{\mathrm{ISO}}(2)$ defined above.
We start from the following unirreps~\cite{Miller1965}
\begin{multline}\label{H4B unirreps 1}
\pi^\lambda_{m\ge n}(\xi_x,\xi_y,\xi_c,\xi_s)=e^{-i(\mathrm{Sgn}(B)n+\lambda+\tilde \delta) \xi_c }e^{-i\xi_s}\left(\frac{n!}{m!}\right)^{\frac{1}{2}}e^{i \mathrm{Sgn}(B)(m-n)\tan^{-1}\left(\frac{\xi_y}{\xi_x}\right)} e^{-\frac{|B| (\xi_x^2+\xi_y^2)}{4}} \\ \left(-i\sqrt{\xi_x^2+\xi_y^2}\left| \frac{B}{2}\right|^{1/2}\right)^{m-n} L_n^{m-n}\left(\frac{|B|}{2}(\xi_x^2+\xi_y^2)\right),
\end{multline}
\begin{multline}\label{H4B unirreps 2}
\pi^\lambda_{m\le n}(\xi_x,\xi_y,\xi_c,\xi_s)=e^{-i(\mathrm{Sgn}(B)n+\lambda+\tilde \delta) \xi_c }e^{-i\xi_s}\left(\frac{m!}{n!}\right)^{\frac{1}{2}}e^{i \mathrm{Sgn}(B)(m-n)\tan^{-1}\left(\frac{\xi_y}{\xi_x}\right)} e^{-\frac{|B| (\xi_x^2+\xi_y^2)}{4}}\\ \left(-i\sqrt{\xi_x^2+\xi_y^2}\left| \frac{B}{2}\right|^{1/2}\right)^{n-m} L_m^{n-m}\left(\frac{|B|}{2}(\xi_x^2+\xi_y^2)\right),
\end{multline}
where $\lambda \in \mathbb{Z}$, $m,n\in\mathbb{N}_0$, $\tilde \delta=1$ if $B>0$ and $\tilde\delta=0$ otherwise,  and $L_n^{m-n}$ are the associated Laguerre polynomials. 
A set of functions in the Hilbert space which transform under these representations can be inferred by comparing the multiplication rule in $\widetilde{\mathrm{ISO}}(2)$ with the group action on the principal bundle $P$. We thus obtain the following basis of functions on $P$:
\begin{multline}
e^{\lambda_0,m}_n |_{m\ge n}(x,y,s)=e^{-i\left(s+h+\frac{B}{2}xy\right)}\left(\frac{n!}{m!}\right)^{\frac{1}{2}}e^{i \mathrm{Sgn}(B)(m-n)\tan^{-1}\left(\frac{y}{x}\right)} e^{-\frac{|B| (\xi_x^2+\xi_y^2)}{4}} \\ \left(-i\sqrt{\xi_x^2+\xi_y^2}\left| \frac{B}{2}\right|^{1/2}\right)^{m-n} L_n^{m-n}\left(\frac{|B|}{2}(\xi_x^2+\xi_y^2)\right),
\end{multline}
\begin{multline}
e^{\lambda_0,m}_n |_{m\le n}(x,y,s)=e^{-i\left(s+h+\frac{B}{2}xy\right)}\left(\frac{m!}{n!}\right)^{\frac{1}{2}}e^{i \mathrm{Sgn}(B)(m-n)\tan^{-1}\left(\frac{y}{x}\right)} e^{-\frac{|B| (x^2+y^2)}{4}}\\ \left(-i\sqrt{x^2+y^2}\left| \frac{B}{2}\right|^{1/2}\right)^{n-m} L_m^{n-m}\left(\frac{|B|}{2}(x^2+y^2)\right).
\end{multline}
where $\lambda_0=-\mathrm{Sgn}(B)-\tilde \delta$. When acted on by the left regular representation of $\widetilde{\mathrm{ISO}}(2)$ these functions transform under the unirrep corresponding to the conjugate of the $\lambda=\lambda_0$ unirrep defined in (\ref{H4B unirreps 1}, \ref{H4B unirreps 2}) above. We know it is sufficient to consider only these unirreps since they satisfy a completeness relation given by
\begin{equation}
\frac{|B|}{2\pi} \sum_{m,n} \left(e^{\lambda_0,m}_n(x^\prime,y^\prime,s^\prime)\right)^*e^{\lambda_0,m}_n(x,y,s)=e^{-i(s-s^\prime)}\delta(x-x^\prime)\delta(y-y^\prime).
\end{equation}
Thus, we can decompose a wavefunction in our Hilbert space into unirreps of $\widetilde{\mathrm{ISO}}(2)$ as
\begin{equation} \label{LandauISO_Decom}
\Psi(x,y,s)=\frac{|B|}{2\pi} \sum_{m,n} e^{\lambda_0,m}_n(x,y,s) f_{m,n},
\end{equation}
where the inverse transform is given by
\begin{equation}
f_{m,n}=\int dx dy (e^{\lambda_0,m}_n(x^\prime,y^\prime,s^\prime))^*\Psi(x,y,s).
\end{equation}
After substituting the decomposition (\ref{LandauISO_Decom}) into the SE, we obtain
\begin{equation}
\frac{|B|}{2\pi}\sum_{m,n}\left( |B|(n+1/2) -E \right) e^{\lambda_0,m}_n(z,ys) f_{m,n}=0.
\end{equation}
Thus, we arrive at the familiar Landau level spectrum
\begin{equation}
\quad E_{m,n}=|B|(n+1/2),\quad \Psi_{m,n}=e^{\lambda_0,m}_n(x,y,s),
\end{equation}
where setting $s=0$ in $e^{\lambda_0,m}_n$ gives us a suitable set of eigenfunctions on $\mathbb{R}^2$.

\subsection{Quantum mechanics on the Heisenberg group} \label{Sec_Heisenberg}
In this Section, we turn to a new example not previously considered in the literature, of particle motion on  the Heisenberg group. We equip $M=\mathrm{Hb}$ with a left-invariant metric, and thus take $G=\mathrm{Hb}$ also. We shall couple the particle to a background magnetic field, corresponding to an $\mathrm{Hb}$-invariant closed 2-form on $\mathrm{Hb}$, for which the magnetic vector potential which appears in the lagrangian shifts by a total derivative under the action of the group $\mathrm{Hb}$ on itself.

While a version of the Heisenberg group appeared in \S \ref{Sec_LandauLevels} (as the central extension of the translation group $\mathbb{R}^2$), for our purposes in this Section we shall redefine the Heisenberg group to be the set of triples $(x,y,z)\in \mathbb{R}^3$ equipped with multiplication law
\begin{equation}\label{Hb group}
(x^\prime,y^\prime,z^\prime)\cdot (x,y,z)=(x+x^\prime,y+y^\prime,z+z^\prime+yx^\prime).
\end{equation}
To avoid any possible confusion, we emphasise that in this Section the Heisenberg group is taken as the original configuration space of our particle dynamics, which we shall reformulate as an equivalent dynamics {\em on a central extension of the Heisenberg group}. This central extension will be a four-dimensional Lie group which we shall denote $\widetilde{\mathrm{Hb}}$.

Before we proceed with writing down the action for this system (and eventually solving for the spectrum using harmonic analysis on $\widetilde{\mathrm{Hb}}$), we first pause to offer a few words of motivation for considering this system, since it does not correspond to any physical quantum mechanics system (although there are indirect links to the anharmonic oscillator, see {\em e.g.} \cite{klink1994nilpotent}). In any case, our motivation is entirely mathematical. Firstly, we wanted a new example where the central extension of Lie groups $0\rightarrow U(1)\rightarrow \tilde G \rightarrow G$ is non-trivial, \emph{i.e.} $\tilde G$ is not just a direct product, and moreover that it corresponds to a non-trivial central extension of Lie algebras $0\rightarrow \mathbb{R}\rightarrow \tilde{\mathfrak g} \rightarrow \mathfrak{g}$. The requirement that a Lie algebra $\mathfrak{g}$ admits a non-trivial central extension requires, by a theorem of Whitehead \cite{whitehead1937certain,whitehead1936decomposition}, that the Lie algebra $\mathfrak{g}$ cannot be semi-simple. Of course, abelian Lie groups provide a source of such non-trivial central extensions, because their Lie algebra cohomology is in a sense maximal (noting that the second Lie algebra cohomology of $\mathfrak{g}$ is isomorphic to the group of inequivalent (up to Lie algebra isomorphisms) central extensions of $\mathfrak{g}$). However, we sought a more interesting example where the original group $G$ is non-abelian. To that end, non-abelian nilpotent Lie groups provide a richer source of suitable central extensions,  because the second Lie algebra cohomology of any nilpotent $\mathfrak{g}$ is at least two-dimensional \cite{DixmierCohomologie}. The Heisenberg Lie algebra, and the corresponding Lie group $\mathrm{Hb}$, provides the simplest such example.

Since we are taking the Heisenberg group to be topologically just $\mathbb{R}^3$, we can cover the target space with a single patch and write the lagrangian using globally-defined coordinates $(x,y,z)$. The action on $\mathrm{Hb}$, including the topological term, is
\begin{equation} \label{Heisenberg_Action}
S=\int dt \left( \frac{1}{2}\left( \dot x^2+ \dot y^2+(\dot z-x \dot y)^2\right) -x \dot z+ \frac{x^2}{2} \dot y\right).
\end{equation}
The kinetic term corresponds to a left-$\mathrm{Hb}$-invariant metric on $\mathrm{Hb}$, as mentioned above, and we have chosen a normalization for the (real-valued) coefficient of the topological term $-x \dot z+ \frac{x^2}{2} \dot y$.\footnote{Note that this is not the most general $\mathrm{Hb}$-invariant topological term we can write down.} This topological term in the lagrangian shifts by a total derivative under the group action (\ref{Hb group}). Following our now-familiar procedure, we thus reformulate the action on a $U(1)$-principal bundle $P$ over $\mathrm{Hb}$, on which $s$ provides a local coordinate in the fibre. The action on $P$ is written
\begin{equation}
S=\int dt \left( \frac{1}{2}\left( \dot x^2+ \dot y^2+(\dot z-x \dot y)^2\right)-\dot s -x \dot z+ \frac{x^2}{2} \dot y\right),
\end{equation}
where the only difference is the $\dot s$ term. By adding this redundant degree of freedom to the action it becomes strictly invariant under the $U(1)$-central extension of $\mathrm{Hb}$ defined by the multiplication law
\begin{equation}\label{CEofH}
(x^\prime,y^\prime,z^\prime,s^\prime) \cdot (x,y,z,s)=\left(x+x^\prime,y+y^\prime,z+z^\prime+yx^\prime,s+s^\prime-zx^\prime-y\frac{x^{\prime 2}}{2}\right),
\end{equation}
which we denote by $\tilde G=\widetilde{\mathrm{Hb}}$.

The total hamiltonian corresponding to the action (\ref{Heisenberg_Action}) is given by
\begin{equation}
H=\frac{1}{2} p_x^2+\frac{1}{2}\left(p_z+x\right)^2+\frac{1}{2}\left(p_y-\frac{x^2}{2}+x\left(p_z+x\right)\right)^2+v(t)\left(p_s+1\right),
\end{equation}
which quantizes to
\begin{multline} \label{Ham_He}
\hat H=-\frac{1}{2} \frac{\partial^2}{\partial x^2} +\frac{1}{2}\left(-i\frac{\partial}{\partial z}+x\right)^2+\frac{1}{2}\left(-i\frac{\partial}{\partial y}-\frac{x^2}{2}+x\left(-i\frac{\partial}{\partial z}+x\right)\right)^2\\+v(t)\left(-i\frac{\partial}{\partial s}+1\right).
\end{multline}
acting on the Hilbert space of square integrable functions on $\widetilde{\mathrm{Hb}}$ that are annihilated by $\left(-i\frac{\partial}{\partial s}+1\right)$.

Because the group $\widetilde{\mathrm{Hb}}$ defined in (\ref{CEofH}) has a nilpotent Lie algebra, its representation theory can be found via Kirillov's orbit method~\cite{kirillov2004lectures}. The unirrep matrix elements that we are interested in,  which in this case are functions on $\widetilde{\mathrm{Hb}}$, are infinite-dimensional, given by
\begin{equation}
\pi^q(r,t;x,y,z,s)=\delta(t-r-x) e^{i(-s+zr+\frac{1}{2} y r^2)+q/2 y},
\end{equation}
which satisfy the completeness relation
\begin{equation}
\int \frac{dq dr dt}{2(2\pi)^2} \left(\pi^q(r,t;x^\prime,y^\prime,z^\prime,s^\prime)\right)^* \pi^q(r,t;x,y,z,s)=e^{-i(s-s^\prime)}\delta(x-x^\prime)\delta(y-y^\prime)\delta(z-z^\prime).
\end{equation}
We thus decompose a wavefunction into unirreps using these functions as our basis elements, $e^{q,t}_r(x,y,z,s)=\pi^q(r,t;x,y,z,s)$, giving us
\begin{equation}
\Psi(x,y,z,s)=\int\frac{dq dr dt}{2(2\pi)^2} e^{q,t}_r(x,y,z,s) f_q(r,t),
\end{equation}
where
\begin{equation}
f_q(r,t)=\int dx^\prime dy^\prime dz^\prime \left( e^{q,t}_r(x^\prime,y^\prime,z^\prime,s^\prime)\right)^*\Psi(x^\prime,y^\prime,z^\prime,s^\prime).
\end{equation}
Using this decomposition, and the expression (\ref{Ham_He}) for the hamiltonian, the SE reduces to
\begin{multline}
-\frac{1}{4(2\pi)^3} \int dq dr dt\ e^{q,t}_r(x,y,z,s) \\ \left(\frac{\partial^2 f_q(r,t)}{\partial t^2}+2E f_q(r,t)-\frac{1}{4}\left((t^2+q)^2+4t^2\right) f_q(r,t)\right)=0.
\end{multline}
The ODE in the parentheses coincides with the SE for an anharmonic oscillator. This differential equation can be solved order-by-order in perturbation theory (in the parameter $q$), as is
discussed in numerous sources, for example~\cite{mcweeny1948quantum}. If the SE of this problem could be solved using other means, this decomposition would allow one to study the eigenstates of the anharmonic oscillator. 

\subsection{Trapped particle in a magnetic field} \label{Sec_varymass}
Our last example will demonstrate our method in a case where the group action $\alpha:G\times M\rightarrow M$ is non-transitive (we saw another such non-transitive example, that of a particle orbiting a dyon, in \S \ref{Sec_dyon}). In particular, we will consider particle dynamics on $M=\mathbb{R}^3$, invariant under the action of a subgroup $G=\mathbb{R}^2\subset \mathbb{R}^3$ corresponding to translations in $x$ and $y$. We will begin this Section by formulating the problem, and introducing the necessary representation theory, to describe a generic such action. We will then consider a special case, in which the components of the inverse metric on $\mathbb{R}^3$ vary quadratically in the $z$ direction. This corresponds, physically, to a $z$-dependent effective mass. In this special case, we shall find that the solutions to the SE become localized (or `trapped') around the $z=0$ plane.

Consider the action
\begin{equation}
S=\int dt \left(\frac{1}{2}\left(a_x(z)\dot x^2+a_y(z) \dot y^2+a_z(z) \dot z^2\right)+V(z)-By \dot x- yf^\prime(z) \dot z\right),
\end{equation}
for a particle moving on $\mathbb{R}^3$. Here $a_x(z)$, $a_y(z)$, $a_z(z)$, $V(z)$, and $f(z)$ are (for now) arbitrary smooth functions of $z$, with $a_x(z)$, $a_y(z)$, and $a_z(z)$ necessarily non-vanishing. This action is quasi-invariant under the non-transitive action of translations in $x$ and $y$, but is not invariant under translations in the $z$ direction. We thus consider an equivalent action on a $U(1)$-principal bundle over $\mathbb{R}^3$, which has to be the trivial one, $P=\mathbb{R}^3\times U(1)$, with coordinates $(x,y,z,s\sim s+2\pi)$. The action is given by
\begin{equation}
S=\int dt \left(\frac{1}{2}\left(a_x(z)\dot x^2+a_y(z) \dot y^2+a_z(z) \dot z^2\right)+V(z)-\dot s-By \dot x- yf^\prime(z) \dot z\right),
\end{equation}
which is strictly invariant under $\tilde G = \mathrm{Hb}$, the Heisenberg group (the unique $U(1)$-central extension of $\mathbb{R}^2$ up to isomorphism), which in this Section we parametrize by $(\zeta_x,\zeta_y,\zeta_s)$, with its group action on the bundle $\mathbb{R}^3\times U(1)$ defined by
\begin{equation}
\tilde \alpha_{(\zeta_x^\prime,\zeta_y^\prime,\zeta_s^\prime)}\circ (x,y,z,s)=(x+\zeta_x^\prime,y+\zeta_y^\prime,z,s+\zeta_s^\prime-\zeta_y^\prime(Bx+f(z))).
\end{equation}

The total hamiltonian corresponding to the above action is given by
\begin{equation}
H=\frac{1}{2 a_x(z)}\left(p_x+By\right)^2+\frac{1}{2a_y(z)}p_y^2+\frac{1}{2a_z(z)}\left(p_z+ y f^\prime(z)\right)^2+V(z)+v(t)(p_s+1),
\end{equation}
which we quantize as the operator
\begin{multline}
\hat H=\frac{1}{2 a_x(z)}\left(-i \frac{\partial}{\partial x}+By\right)^2-\frac{1}{2a_y(z)} \frac{\partial^2}{\partial y^2}+\frac{1}{2a_z(z)}\left(-i \frac{\partial}{\partial z}+ y f^\prime(z)\right)^2+V(z)\\+v(t)\left(-i\frac{\partial}{\partial s}+1\right).
\end{multline}
We decompose a wavefunction into unirreps of $\mathrm{Hb}$, exactly as in \S \ref{Sec_LandauLevels}. The difference in this non-transitive case is that the coefficients of the unirreps will depend on $z$, \emph{viz.}
\begin{equation}\label{NonTran_Deco}
\Psi(x,y,z,s)=\frac{2\pi}{|B|} \int dr dt e^{B,t}_r( x,y,s)f(r,t;z),
\end{equation}
where as before
\begin{equation}
e^{B,t}_r( x,y,s)=e^{iBxr-is}\delta(r+y-t).
\end{equation}
This however, now transforms under the unirrep of $\mathrm{Hb}$ defined by
\begin{equation} \label{Hb transitive unirrep}
\tilde \pi^{-B}(r,t;\zeta_x,\zeta_y,\zeta_z)=\left(\exp\left(if(z) \zeta_y\right) e^{B,t}_r( \zeta_x,\zeta_y,\zeta_s)\right)^*,
\end{equation}
which takes account of the transformation of $s$ which is not the same as $\zeta_s$, as was the case in our previous examples.
This can be seen from
\begin{equation}
\begin{aligned}
\rho((\zeta_x^\prime,\zeta_y^\prime, \zeta_s^\prime))\cdot e^{i(Bxr-s)}\delta(r+y-t)=e^{i(B(x-\zeta_x^\prime)r-i(s-(\zeta_s^\prime+B\zeta_y^\prime\zeta_x^\prime)+\zeta_y^\prime(Bx+f^\prime(z))}\delta(r+y-\zeta_y^\prime-t),\\
=\int dq \left( e^{if^\prime(z)\zeta_y} e^{i(B\zeta_x q-\zeta_s)} \delta(q+\zeta_y-r)\right)^*e^{i(Bxq-s)}\delta(q+y-t).
\end{aligned}
\end{equation}
Upon this decomposition, the SE reduces to the following PDE 
\begin{equation}
\left(\frac{B^2t^2}{2a_x(z)}-\frac{\partial_t^2}{2a_y(z)}+\frac{\left(-i\partial_z +(t-r)f^\prime(z)\right)^2}{2a_z(z)}+V(z)\right)  f(r,t;z)=E  f(r,t;z).
\end{equation}
Even in this case where $G$ acts non-transitively on $M$, we see that using harmonic analysis (on a central extension) has removed derivatives with respect to the two variables $x$ and $y$, and replaced them with derivatives with respect to the single variable $t$, which labels distinct copies of the unirrep (\ref{Hb transitive unirrep}) that appears in the Hilbert space.

As a specific example where this PDE can be solved analytically, we take $f^\prime(z)=0$, $V(z)=0$, $a_z(z)=1$, and $a_x(z)=a_y(z)=(a+z^2)^{-1}$ with $a\in\mathbb{R}_+$. That is, we do not consider the addition of a $z$-dependent potential, but we do consider a (specific) $z$-dependent metric on $\mathbb{R}^3$. This equation admits solutions by separation of variables, {\em viz.} $f(r,t;z)=f(r,t)g(z)$, after which $f(r,t)$ is found to satisfy a simple harmonic oscillator equation (with quantum number $n\in \mathbb{Z}$) analogous to (\ref{LandauODE}). Likewise, $g(z)$ is then found to satisfy
\begin{equation}
\left(-\frac{1}{2} \frac{\partial^2}{\partial z^2} g(z)+|B|(n+1/2) g(z)(a+z^2)\right)=Eg(z), \quad n\in \mathbb{Z},
\end{equation}
which is simply the harmonic oscillator equation again. As such the $z$-dependence may be written in the form
\begin{equation}
g(z)=H_m\left(\left(|B|(2n+1)\right)^{1/4} z\right) e^{-\sqrt{|B|(2n+1)} z^2/2}, \quad \quad m\in \mathbb{Z}.
\end{equation}
We can obtain an expression for the eigenstates by inverting the decomposition in (\ref{NonTran_Deco}) and setting $s=0$, to obtain functions on $\mathbb{R}^3$. Following a similar procedure to that in \S \ref{Sec_LandauLevels}, we arrive at the eigenstates
\begin{multline}
\Psi_{m,n,\alpha}(x,y,z)=H_m\left(\left(|B|(2n+1)\right)^{1/4} z\right) e^{-\sqrt{|B|(2n+1)} z^2/2} e^{i\alpha x} \\ H_n(\sqrt{|B|}(y+\alpha/B)) e^{-\frac{|B|}{2} (y+\alpha/B)^2},
\end{multline}
where $\alpha\in \mathbb{R}$.
The energy levels depend only on the two quantum numbers $n$ and $m$, both in $\mathbb{Z}$, and are given by
\begin{equation}
E_{m,n,\alpha}=\sqrt{|B|(2n+1)}(m+1/2)+a|B|(n+1/2).
\end{equation}
Thus, interestingly, the eigenstates for this system appear to be trapped in the $z$-direction (even though na\"{i}vely one may expect the opposite).

\section{Symmetry reduction in magnetic backgrounds} \label{sec_nonTransv}
Back in \S \ref{Formalism}, we claimed that a certain condition (\ref{Manton condition}) on the field strength 2-form $\omega$, which we called the Manton condition, must be satisfied in order for particle motion in that magnetic background to result in a $G$-invariant quantum mechanics. Specifically, this condition, which was proven (in the context of sigma models in any dimension) in ~\cite{Davighi2018}, demands that the contraction of $\omega$ with each vector field generating the $G$ action on $M$ must be an exact 1-form. In all the examples considered so far in this paper, that condition has been satisfied, and thus, while there might not necessarily have existed a $G$-invariant lagrangian corresponding to that topological term, we saw that there nevertheless always existed a $G$-invariant {\em action}.

When the Manton condition is violated, however, there will exist non-contractible wordlines in $M$ on which a $G$-invariant action cannot be written down at all (the necessity of non-contractible cycles in $M$ for the Manton condition to fail makes manifest the topological character of this condition). In that sense, the symmetry group of a particle on $M$ in the presence of such a Manton condition-violating magnetic background is reduced from $G$ down to some subgroup $K \subset G$ on which the Manton condition holds, which one may determine.\footnote{In \cite{Davighi:2018xwn}, we discussed a number of analogue examples from field theory in which the Manton condition is violated in a similar way, namely in four-dimensional Composite Higgs models (in which the r\^ole of the magnetic background is replaced by a Wess-Zumino term). In these examples, and indeed for sigma models in any number of dimensions ({\em i.e.} not just the $(0+1)$-dimensional version that is the subject of the present paper), the result of violating the Manton condition is the same; namely, there is a reduction in the symmetries of the quantum system.}
 Since the classical equations of motion nevertheless retain invariance under all of $G$, this symmetry breaking due to the magnetic background may be interpreted as an anomaly of the quantum theory, albeit of a kind that might be unfamiliar to many readers. In particular, this kind of anomaly does not derive from an inability to appropriately regularize the path integral measure for fermions in a way that is compatible with the symmetry; indeed, this anomaly is not related to fermions at all, but follows only from topological considerations. Furthermore, the {\em lagrangian} may still shift by a total derivative under $K$, in which case we should pursue a similar strategy as in the rest of this paper and write an equivalent dynamics which is invariant under a $U(1)$ central extension $\tilde K$ of $K$.

In this Section, we elucidate in more detail how this type of anomaly can arise, by discussing two examples. Firstly, we review quantum mechanics on a torus, which was discussed in~\cite{Davighi2018} (in fact, this example was considered by Manton ~\cite{Manton1983,Manton1985}, where this type of anomaly was first observed). We then turn to a new example where the Manton condition is violated, which is quantum mechanics on the compact Heisenberg manifold. In both cases, it is not our goal in this Section to actually solve for the spectrum of these systems using harmonic analysis; rather, here, we content ourselves with a careful analysis of the symmetries that are preserved in the quantum theory, {\em i.e.} with the determination of the unbroken subgroup $K$ in both examples.

\subsection{Quantum mechanics on the torus}

We start with the simpler example of quantum mechanics on the $2$-torus~\cite{Manton1983,Manton1985}, $M=(\mathbb{R}/\mathbb{Z})^2$, parametrized by two periodic coordinates $x\sim x+1$ and $y\sim y+1$, with translation symmetry $G=U(1)\times U(1)$. We define a magnetic background corresponding to the translation invariant field strength 2-form $\omega=2\pi B dx \wedge dy$, for $B \in \mathbb{Z}$ (where this quantization condition on $B$ ensures that $\omega$ is the curvature of a well-defined $U(1)$-principal bundle over $T^2$, for which the first Chern class must of course be an integer). However, contracting this 2-form with the vector field  generating translations, $X=a_x \partial_x + a_y \partial_y$, yields
\begin{equation}
\iota_{a_x \partial_x + a_y \partial_y} \left(2\pi B dx \wedge dy\right) = 2\pi a_x B dy - 2\pi a_y B dx,
\end{equation}
which is a closed but not an exact 1-form on $T^2$ and thus violates the Manton condition (unless $a_x = a_y = 0$ or $B=0$).

To see that one cannot indeed write down a $G$-invariant action (or, more precisely, action phase),  
consider a loop $\gamma$ on the torus at constant $x=x_0$ which wraps around the $y$-direction. On such a loop, we may introduce the vector potential $A=2\pi B x dy$ such that $\omega=dA$, and from here evaluate the action phase, which is the holonomy over this loop. It is here sufficient to integrate $A$ over $\gamma$, yielding the action phase $e^{i2\pi B x_0 }$. Note that the value of the holonomy of a connection (evaluated over a given loop) only depends on the curvature $\omega$ and on its characteristic class, which may contain torsion information. Thus, the action phase that we evaluate does not depend on our particular choice of $A$, for fixed $\omega$ and characteristic class.

This is sure enough not invariant under generic translations in the $x$ direction, but only under discrete translations $x\rightarrow x+a/B$ for $a \in \mathbb{Z}_B$. Similarly, we may conclude (from evaluating the holonomy over a loop in the $x$ direction at constant $y$) that the action phase is only invariant under discrete translations in the $y$ direction also, $y\rightarrow y+a/B$ for $a \in \mathbb{Z}_B$. Thus, the symmetry group is here reduced from $G=U(1) \times U(1)$ to the discrete group $K=\mathbb{Z}_B \times \mathbb{Z}_B$. This fact was first derived by explicitly solving the SE for this system, and finding that the corresponding eigenfunctions do not respect the continuous translation invariance of the classical equations of motion. Rather, the eigenfunctions of the hamiltonian become localized when the magnetic field is switched on, preserving only the discrete  $\mathbb{Z}_B \times \mathbb{Z}_B$ symmetry ~\cite{Manton1985}.

\subsection{Quantum mechanics on the compact Heisenberg manifold}

Our second example of this type of anomaly is new, and is that of quantum mechanics on the Heisenberg manifold. The Heisenberg {\em manifold}, to be contrasted with the Heisenberg {\em group} discussed in \S\S \ref{Sec_LandauLevels} and \ref{Sec_Heisenberg}, is defined by quotienting the (continuous) Heisenberg group (\ref{Hb group}) by its discrete subgroup in which $x$, $y$, and $z$ are all integers. Thus, the Heisenberg manifold, which we can denote by the coset space $M=\mathrm{Hb}(\mathbb{R})/\mathrm{Hb}(\mathbb{Z})$, is parametrized by $(x,y,z) \in \mathbb{R}^3$ with the equivalence relation
\begin{eqnarray}\label{Hb equ}
x\sim x+p,\\
y\sim y+m,\\
z\sim z+n+xm,\label{Hb equ z}
\end{eqnarray}
where $(p,n,m) \in \mathbb{Z}^3$. 
We shall consider quantum mechanics on this space in the presence of a magnetic background, with symmetry group $G=\mathrm{Hb}(\mathbb{R})$, which acts on $[(x,y,z)]\in M$ by left translation (\ref{Hb group}).

In particular, we consider a topological term in the action for which the curvature 2-form is
\begin{equation}\label{Hb mfd omega}
\omega=B dx \wedge dy, \quad B \in \mathbb{Z},
\end{equation}
which is the unique topological term on $M=\mathrm{Hb}(\mathbb{R})/\mathrm{Hb}(\mathbb{Z})$, as it is the only closed left-invariant 2-form on $\mathrm{Hb}$ which is constant on the equivalence classes defined in (\ref{Hb equ}-\ref{Hb equ z}).
The quantization condition on the coefficient $B$ ensures that $\omega$ is an integral 2-form on $M$ (meaning its integral over any 2-cycle in $M$ evaluates to an integer), and thus the $U(1)$-principal bundle over $M$, which defines the background magnetic field, is well-defined.

Despite being invariant under the action of $G=\mathrm{Hb}$, the 2-form $\omega$ does not, however, satisfy the (stronger) Manton condition. In our coordinates, a basis for the
right-$\mathrm{Hb}$-invariant vector fields (which generate {\em left} translations on $M$) is
\begin{equation}
\{X_1, X_2, X_3\}=\{\partial_x+y\partial_z,\ \partial_y,\ \partial_z\}. \label{RIVFs on H}
\end{equation}
When a linear combination of these vector fields is contracted with $\omega$, we obtain
\begin{equation}
\iota_{\alpha_1 X_1+ \alpha_2 X_2+ \alpha_3 X_3}(B dx \wedge dy) =B \left( \alpha_1 dy - \alpha_2 dx \right).
\end{equation}
Just as the 1-form $d\theta$ on a circle is closed but not exact because $\theta\sim \theta+2\pi$, so $dx$ and $dy$ are closed but not exact 1-forms on the Heisenberg manifold because of the identifications in (\ref{Hb equ}-\ref{Hb equ z}). Thus, the Manton condition is only satisfied for $X_3$, hence the topological term remains invariant on the 1-parameter subgroup that corresponds to the integral curves of $X_3$. Indeed, it is not surprising that the Manton condition is satisfied for $X_3$, but not for $X_1$ or $X_2$, because it was proven in ~\cite{Davighi2018} that the Manton condition is necessarily satisfied for any element in $[\mathfrak g, \mathfrak g]$, which in this case is just $X_3$.

Nonetheless, the continuous symmetries that are generated by $X_1$ and $X_2$ are not broken completely; as in the case of quantum mechanics on the torus discussed above, a discrete subgroup of the $\mathbb{R}^2$ subgroup generated by $X_1$ and $X_2$ remains unbroken.
The unbroken symmetry group $K$ turns out to be the subgroup
\begin{equation}\label{unbroken K group}
K=\left\{\left( \frac{n}{B},\frac{m}{B},b \right) \in \mathrm{Hb}\ |\ b\in\mathbb{R},\ (n,m)\in \mathbb{Z}_B\times \mathbb{Z}_B \right\}.
\end{equation}
This group is a (non-trivial) central extension (by $\mathbb{R}$) of the discrete subgroup $\mathbb{Z}_B\times \mathbb{Z}_B$, defined by the exact sequence
\begin{equation}
\begin{tikzcd}
0\arrow{r}{}& \mathbb{R} \arrow{r}{}& K\arrow{r}{} &\mathbb{Z}_B\times \mathbb{Z}_B\arrow{r}{} &0,
\end{tikzcd}
\end{equation}
where the group homomorphisms involved should be obvious given (\ref{unbroken K group}).
The lagrangian, including both the kinetic energy and this topological term, is in this case strictly invariant under this subgroup $K$, so there is no need to pass to a $U(1)$-central extension.

\section{Discussion} \label{Sec_Discussion}

We have formulated the quantum mechanics of a particle moving on a manifold $M$, with dynamics invariant under the action of a Lie group $G$, in the presence of a background magnetic field. The coupling to a magnetic background, which is included via a topological term in the action, defines a $U(1)$-principal bundle $P$ over $M$ with connection. We suggest that such a dynamics should be recast using an equivalent action on this principal bundle $P$, for two reasons. Firstly, a globally-defined lagrangian is guaranteed to exist only on $P$, but not on $M$ itself. Secondly, even if a lagrangian were to be defined (locally) on $M$, this lagrangian would not in general be invariant under the action of $G$; rather, due to the presence of the topological term, it might shift by a total derivative. Once reformulated on $P$, we have shown that the lagrangian will be strictly invariant, not under $G$, but under a larger symmetry group $\tilde G$, which is a $U(1)$-central extension of $G$. We show how to construct this central extension $\tilde G$, which is a {\em bona fide} symmetry group of the system, in the general case.

We have discussed a plethora of examples in which these two (related) complications arise in coupling a particle to a magnetic background, and in every case show explicitly how reformulating the dynamics on the principal bundle $P$ remedies the issues. To highlight just one example, we have revisited the seemingly humble problem of quantizing a rotating rigid body in three dimensions, a system that is familiar from every undergraduate quantum mechanics course, which is equivalent to particle motion on the configuration space $SO(3)$. What is perhaps less familiar, and which is of interest to us in this paper, is that there is in fact a topological term in this theory. This topological term, whose existence stems from the non-vanishing  cohomology group $H^2(SO(3),\mathbb{Z})\cong \mathbb{Z}/2$, can only be written as a globally-defined term in the lagrangian if we pass to a principal bundle over $SO(3)$. There are two choices of such bundle, both of which are isomorphic to central extensions of $SO(3)$; the bundle is either $U(2)$, or $SO(3)\times U(1)$. We show that the former choice corresponds to a term in the action phase that evaluates to $-1$ upon traversing closed loops in the configuration space, and thus has the affect of ascribing fermionic character to the rigid body.

The second main feature of this paper is the introduction of a new method for {\em solving} the Schr\"odinger equation for such quantum mechanical systems with magnetic backgrounds. Our method exploits the group-theoretic structure of the problem, by decomposing the Hilbert space into unitary irreducible representations of the central extension $\tilde G$. The method is thus very general; indeed, we show that it is a suitable match for the generality of the problem which we are attempting to solve.
Because the Hilbert space carries a {\em bona fide} representation of the group $\tilde G$ (but not the group $G$, in which the Hilbert space carries only a projective representation), we expect that such a decomposition should yield a solution for the spectrum of the corresponding hamiltonian. In the example of the fermionic rigid body mentioned above, we immediately see the appearance of spin-$\frac{1}{2}$ representations in the spectrum by decomposing into representations of $\tilde G=U(2)$, thus exhibiting the non-trivial connection between topological terms in the action and representation theory.

We proceed to illustrate in all our examples how methods from harmonic analysis can be used to decompose the Hilbert space into representations of a central extension $\tilde G$, and in all cases this decomposition is found to be fruitful, typically reducing the SE to a family of ODEs whose solutions might be known. Our chosen examples range over some much-loved problems in quantum mechanics, including that of a particle moving on a plane in a uniform perpendicular magnetic field, a charged particle moving in the field of a magnetic monopole, and a charged particle moving in the field of a dyon. This last example illustrates the virtues of our method even in cases where the group $G$ acts non-transitively on $M$, in reducing the problem to one on the space of orbits of $G$. We also study some new examples, including a particle moving on the Heisenberg group in the presence of a magnetic background, for which the Schr\"odinger equation is found to reduce, after decomposing into irreducible representations of a central extension of the Heisenberg group, to that of an anharmonic oscillator.

We anticipate that there are many more quantum mechanics problems which can be described by dynamics on a manifold with invariance under a Lie group action, and a coupling to a magnetic field, because this setup is a very general one. For example, the cases where $M=\mathbb{R}^n$ or $SO(n)$ appear ubiquitously in physics and chemistry, and one might describe more realistic molecular systems moving in magnetic fields, for example, by using a perturbative analysis around these simple cases. Another possible source of examples, of interest to condensed matter physicists and particle theorists, might be provided by quantum field theories admitting instanton solutions, in which great insight can be gained by solving for quantum mechanics on the instanton moduli space. Since such theories typically also contain topological terms in the action, the method of solution we have outlined in this paper, in which we first construct the {\em bona fide} symmetry group using central extensions, and then bring to bear the heavy machinery of harmonic analysis, would be applicable.

Finally, we observe that all
the quantum mechanical problems studied in this paper have had topological terms that are linear in  time derivatives. This is not, however, the only possibility for lagrangians which are quasi-invariant under the action of a symmetry Lie group $G$. For an example where this is not the case, consider a free non-relativistic particle. This can be described in terms of motion in space which has a transitive action by the Galileo group, but is such that the lagrangian is not invariant, but shifts by a total derivative under a boost. It turns out that the familiar kinetic term for such a non-relativistic free particle, {\em  viz.} $\frac{1}{2}m\dot x ^2$, which is {\em quadratic} in time derivatives rather than linear, is nonetheless the result of a topological term in the action. To formulate and solve this example using the methods employed here requires the use of so-called ‘inverse Higgs constraints’. These constraints are equivalent to the removal of Goldstone bosons by the equations of motion, and they add complications to the methods introduced in this paper; in particular, once the inverse Higgs constraint is applied we can no longer na\"ively rewrite the topological term as the holonomy of a connection on a principal bundle.
This, and the other complications that arise in such cases, will be addressed in a future work.

\section*{Acknowledgments}
We thank Nakarin Lohitsiri for helpful discussions. 
JD is supported by The Cambridge Trust and STFC consolidated grant ST/P000681/1. BG is partially supported by STFC consolidated grant ST/P000681/1 and King's College, Cambridge. JTS is supported by STFC consolidated grant ST/S505316/1.

\appendix
\section{Mathematical prerequisites} \label{sec_useful_mathematics}
In this Appendix we will present, through an example, a brief summary of some of the mathematical concepts used in this paper. A more detailed discussion is given in \emph{e.g.} \cite{nakahara2003geometry,kobayashi1963,Naber2011}, which are the main references for this Appendix.

We start by defining a fibre bundle, using as our prototype the (principal) fibre bundle introduced in \S\ref{Sec_DiracMonopole} to describe the magnetic monopole with unit charge. A fibre bundle consists of a pair of smooth manifolds, $P$ the \emph{total space} and $M$ the \emph{base space}, and a surjective map $\pi:P \rightarrow M$ between them called the \emph{projection}. In our example the total space is $P=S^3$, which can be embedded in $\mathbb{C}^2$ using the parametrization $(z_1=\cos(\theta/2)e^{i(\chi+\phi)/2},z_2=\sin(\theta/2)e^{i(\chi-\phi)/2} )\in \mathbb{C}^2$, where $\theta \in[0,\pi]$, $\phi \in[0,2\pi)$, and $\chi \in [0,4\pi)$. The base space is here $M=S^2$, which we embed in $\mathbb{R}^3$, with the projection map $\pi:S^3\rightarrow S^2$ defined by
$\pi\left(z_1,z_2\right)=\left(\sin(\theta)\cos(\phi), \sin(\theta)\sin(\phi),\cos(\theta)\right)$. 
The pre-image $\pi^{-1}(m)$, for any point $m\in M$, is diffeomorphic to the same differential manifold, $F$, known as the \emph{typical fibre} of the fibre bundle. For the bundle $\pi:S^3\rightarrow S^2$ the typical fibre is $F=S^1$, as can be seen from
$\pi^{-1}\left((0,0,0)\right)=\{( e^{i\frac{\chi}{2}}, e^{i \frac{\chi}{2}})\mid \chi \in [0,4\pi)\}$, for example.

The base space $M$ is equipped with an open covering $\{U_i\}$ and a collection, $\{\phi_i\}$, of \emph{local trivialisations}. Local trivialisations are diffeomorphisms of the form $\phi_i:\pi^{-1}(U_i)\rightarrow U_i \times F$ with $\pi\circ \phi_i(m,f)=m$ for all $m \in U_i$.  On double intersections of open sets, there are \emph{transition functions}, $t_{ij}: U_i \cap U_j \rightarrow G$, from $M$ to some group $G$, known as the \emph{structure group}. There is a left-action of the group $G$ on the fibre $F$  defined such that $\phi_j^{-1}(m,f)=\phi_i^{-1}(m,t_{ij}(m)f)$. In the context of our example $\pi:S^3\rightarrow S^2$, an open covering of $S^2$ is given by the charts $\{U_+,U_-\}$ defined in \S \ref{Sec_DiracMonopole}, and a valid possible set of local trivialisations over this covering is
$\phi_+(z_1,z_2)=\left((\theta,\phi), e^{i(\chi+\phi)/2}\right)$ and
$\phi_-(z_1,z_2)=\left((\theta,\phi), e^{i(\chi-\phi)/2}\right)$.
The structure group is $U(1)$ with a single transition function given by $t_{+-}(\theta,\phi)=e^{-i\phi}$. The inverse of these trivialisations are given by 
$\phi_+^{-1}((\theta,\phi), e^{is_+})=\left(\cos(\theta/2) e^{i s_+}, \sin(\theta/2) e^{i(s_+-\phi)}\right)$ and
$\phi_-^{-1}((\theta,\phi), e^{is_-})=\left(\cos(\theta/2) e^{i (s_-+\phi)}, \sin(\theta/2) e^{is_-}\right)$.

In this work we make frequent use of a specific type of fibre bundle, known as a principal (fibre) bundle. In a principal bundle, the structure group $G$ is a Lie group which, as a manifold, is diffeomorphic to the typical fibre $F$. In addition, the Lie group $G$ has a right action, denote it $R_g$, on $P$  such that $\pi\circ R_g=\pi$, and that acts both freely and transitively on each fibre.  For our example, $G=U(1)$ which is diffeomorphic to $S^1$ as a manifold, and we can define a suitable right action $R_g$, for $g=e^{i\delta} \in U(1)$, by $R_{e^{i\delta} } \phi_\pm^{-1}((\theta,\phi),e^{is}) =\phi_\pm^{-1}((\theta,\phi),e^{is+i\delta})$,
which is both free and transitive.

Next, we define the concept of a \emph{local section}, $\sigma_i$, which is a smooth map $\sigma_i:U_i\rightarrow P$ such that $\pi\circ \sigma_i=\mathrm{id}_M$. In this paper we have at times described wavefunctions as sections on the hermitian line bundle associated with the $U(1)$-principal bundle $P$. This refers to a set of functions, $s_{i}:U_i\rightarrow \mathbb{C}$, defined for each open set $U_i$ in our cover, which on double intersections are related by $s_j=t_{ij}  s_i$, where $t_{ij}$ are the $U(1)$-valued transition functions of the principal bundle~$P$.

On a principal bundle, $\pi:P\rightarrow M$, we can define a \emph{principal-connection $1$-form} (or simply a \emph{connection} for short). This is a $1$-form on $P$ with value in the Lie algebra, $\mathfrak{g}$, of the Lie group $G$. A connection must also satisfy the following conditions
\begin{equation}
\begin{aligned}
A(X^\#)&=X,\\
R_g^*A&=\mathrm{Ad}_{g^{-1}} A,
\end{aligned}
\end{equation}
where $X$ is in the Lie algebra $\mathfrak{g}$, and the vector field $X^{\#}$ on $P$ is defined by
\begin{equation}
X^\# f(p)=\left.\frac{d}{dt} f\left(R_{e^{itX}}\cdot p\right) \right|_{t=0}
\end{equation}
for $p\in P$ and $f:P\rightarrow \mathbb{R}$.

On the principal bundle $\pi:S^3\rightarrow S^2$, the $1$-form $A= d\chi/2+ \cos\theta\ d\phi/2$ can be seen to be a valid connection as follows.
Firstly, it is $\mathbb{R}$-valued which is as required since the Lie algebra of $U(1)$ is $\mathbb{R}$. Secondly, from the right action of $e^{it\delta} \in U(1)$ on $P$ we can deduce that the vector field $X^\#=2\delta \frac{\partial}{\partial \chi}$, which implies $A(X^\#)=\delta=X\in\mathfrak{g}$ as required. Lastly, it can be seen that both terms of $A$ are invariant under $R_g^*$, meaning the second condition is satisfied since $\mathrm{Ad}_{g^{-1}} A=A$ for $U(1)$.

Throughout this paper we will often resort to using local expressions for the connection, which can be obtained using a corresponding pair of sections and trivialisations. Notably, given local sections $\sigma_i$, the corresponding trivialisation, known as the \emph{canonical local trivialisation}, is defined by 
\begin{equation}
\phi_i(p)=(\pi(p),g_i),
\end{equation} 
where $p\in \pi^{-1}(U_i)$ and $g_i$ are related by $p=R_{g_i} \sigma_i(\pi(p))$.
Given this, and letting $A_i=\sigma_i^* A$, locally
\begin{equation}
\left.A\right|_{U_i}=g_i^{-1}\pi^* A_i g_i-ig_i^{-1} dg_i,
\end{equation}
where $d$ is the exterior derivative on $P$.  Equivalently, and going the other way, sections may be defined from a given choice of local trivialisation.

It turns out that the trivialisation defined above for our example is the canonical local trivialisation that corresponds to the pair of sections 
$\sigma_+(\theta,\phi)=\left(\cos(\theta/2) , \sin(\theta/2)e^{-i\phi}\right)$ and
$\sigma_-(\theta,\phi)=\left(\cos(\theta/2)e^{i\phi}, \sin(\theta/2)\right)$,
which can be seen by simply setting $s_+$ and $s_-$ to zero in the formulae for $\phi_{\pm}^{-1}$. Then $A_+=\frac{1}{2}(-1+\cos(\theta)) d\phi$ and $A_-=\frac{1}{2}(1+\cos(\theta))d\phi$. Furthermore we have that $g_+=e^{is_+}$ and $g_-=e^{is_-}$, which gives us the local expressions for the connection,
$\left.A\right|_{U_+}=ds_++\frac{1}{2} (-1+\cos(\theta)) d\phi$ and
$\left.A\right|_{U_-}=ds_-+\frac{1}{2}(1+\cos(\theta))d\phi$.

Finally, we must introduce the concepts of holonomy and horizontal lift. Given a connection  we can define the \emph{horizontal lift of a curve} $\gamma(t)$ in $M$ as a curve $\gamma_{hl}(t)$ in $P$ such that $\gamma(t)=\pi(\gamma_{hl}(t))$, and such that the tangent vector at each point, call it $Y_{\gamma_{hl}(t)}$, satisfies $A(Y_{\gamma_{hl}(t)})=0$, \emph{i.e.} is horizontal with respect to the connection. The horizontal lift of a curve is unique, up to specifying the start point in the fibre above, say, $\gamma(0)$. 
As an example, given our above connection $A= d\chi/2+ \cos\theta\ d\phi/2$, the horizontal lift of the curve $\gamma(t)=(\cos t,\sin t,0)$  in $S^2$, starting at $(z_1=1,z_2=0)\in S^3$, is given simply by
$\gamma_{hl}(t)=(1,0)$,
which has the horizontal tangent vector
$Y_{\gamma_{hl}(t)}=\frac{\partial}{\partial \phi}-\frac{\partial}{\partial \chi}$.

Using a horizontal lift we can define the holonomy. The \emph{holonomy of a loop} $\gamma(t)$ in $M$ for $t\in[0,2\pi]$ is defined as the element $g\in G$ such that
\begin{equation}\label{hol 1}
\gamma_{hl}(2\pi)=R_g \gamma_{hl}(0).
\end{equation}
For the specific $\gamma_{hl}$ in our example the holonomy is trivially $1$, because $\gamma_{hl}(2\pi)= \gamma_{hl}(0)$.
We can also derive an equivalent (and perhaps more familiar) formula for the holonomy which involves integrating the connection $A$. To wit, let $\tilde \gamma(t)$  be a loop in $P$ which projects down to $\gamma(t)$ under $\pi$. For any such loop $\tilde \gamma(t)$, the horizontal lift is related to $\tilde \gamma(t)$ by
 \begin{equation}\label{hol 2}
 \gamma_{hl}(t)=R_{\left(e^{-i \int_0^t \tilde \gamma^* A}\right)} \tilde \gamma(t).
 \end{equation}
Using (\ref{hol 1}) and (\ref{hol 2}), one finds that the holonomy of $\gamma(t)$ (with respect to the connection $A$) is equal to $e^{-i \int_0^{2\pi} \tilde \gamma^* A}$. In our example, $\gamma_{hl}(t)$ is a already a loop and thus, again, it is obvious that the holonomy is $1$.

\section{Rudiments of harmonic analysis with constraints}\label{Rudiments_of_HA}
In this Appendix we will review, by way of an example, the form of harmonic analysis used throughout this paper. 
The example we will use is that of planar motion in a magnetic field, as discussed in \S\ref{Sec_LandauLevels}.

In all the examples in this paper, we decompose the left-regular representation of $\tilde G$, which recall is a central extension by $U(1)$ of the original group $G$ (constructed in \S \ref{Formalism}), into unirreps of $\tilde G$. In our prototypical example, we have $G=M=\mathbb{R}^2$ and $\tilde G=\mathrm{Hb}$, and the left-regular representation of $\mathrm{Hb}$ is defined by
\begin{equation} \label{landau_left_Regular}
\rho((x^\prime,y^\prime,s^\prime))\cdot \Psi(x,y,s)=\Psi(x-x^\prime,y-y^\prime,s-s^\prime-Bx^\prime y^\prime+By^\prime x).
\end{equation}
for $\Psi(x,y,s)\in\mathcal{H}$, where the Hilbert space $\mathcal{H}$ was defined in (\ref{landau hilbert space}).

In this example we first decompose a general $\tilde \Psi(x,y,s)\in L^2(\mathrm{Hb})$ into unirreps of $\mathrm{Hb}$, following~\cite{Gripaios2016}:
\begin{equation} \label{full_decomposition}
\tilde{\Psi}(x,y,s)=\sum_k \int dr dt \frac{|k|}{4\pi^2}D^k(r,t;x,y,s)g^k(r,t) \in L^2(\mathrm{Hb}),
\end{equation}
where recall the unirreps $D^k$ are
\begin{equation}
D^k(r,t;x,y,s)=e^{ik(xr-s/B)} \delta(r+y-t), \qquad k/B \in \mathbb{Z},
\end{equation}
which transform under the left-regular representation as
\begin{equation}
\rho((x^\prime,y^\prime,s^\prime))\cdot D^B(q,t;x,y,s)=\int D^{-B}(q,r;x^\prime,y^\prime,s^\prime)D^B(q,t;x,y,s) dq,
\end{equation}
\emph{i.e.} in the unirrep $D^{-B}$. inverse transform is
\begin{equation}
g^k(r,t)=\int dx dy ds \left(D^k(r,t;x,y,s)\right)^*\Psi(x,y,s).
\end{equation}
These unirreps satisfy the Schur orthogonality relation 
\begin{equation}\label{orthog}
\int dx dy ds \left(D^k(r,t;x,y,s)\right)^*D^{k^\prime}(r^\prime,t^\prime;x,y,s)=\frac{4\pi^2}{|k|} \delta_{\frac{k}{B}, \frac{k^\prime}{B}}\delta(r-r^\prime) \delta(t-t^\prime).
\end{equation}
Enforcing the constraint $(-i\partial_s +1)\tilde\Psi = 0$, and using the orthogonality relation (\ref{orthog}), immediately implies $g^k(r,t)=0,\ \forall k\neq B$. 
We can then write
\begin{equation} \label{Landa_Decomp2}
\Psi(x,y,s)= \int dr dt \frac{|B|}{2\pi}D^B(r,t; x,y,s) f(r,t) \in \mathcal{H},
\end{equation}
thus recovering the decomposition in (\ref{Landa_Decomp}),
where $g^k(r,t)=2\pi \delta_{\frac{k}{B},1} f(r,t)$, and
the inverse of this decomposition is given by
\begin{equation} \label{Landa_Decomp_Inverse}
f(r,t)=\int dx^\prime dy^\prime \left( D^B(r,t; x^\prime,y^\prime,s^\prime)\right)^* \Psi(x^\prime,y^\prime,s^\prime).
\end{equation}
In other words, we may restrict our decomposition to those unirreps which satisfy the constraint. This restricted subspace of unirreps (which satisfy the constraint) inherits the following completeness relation
\begin{equation} \label{restricted unirreps}
\int dr dt \frac{|B|}{2\pi} \left(D^B(r,t;x^\prime,y^\prime,s^\prime)\right)^* D^B(r,t;x,y,s)=e^{-i(s-s^\prime)}\delta(x-x^\prime)\delta(y-y^\prime).
\end{equation}
It seems plausible that, under suitably general assumptions, one may decompose a general state $\Psi \in \mathcal{H}$ into a basis of unirreps of $\tilde G$ which satisfy the constraint, following a similar procedure to that used in this example. We have indeed found this to be the case in all examples considered, as can be verified on a case-by-case basis by obtaining a completeness relation on the Hilbert space $\mathcal{H}$, analogous to (\ref{restricted unirreps}).

\bibliography{references}
\end{document}